\documentclass[12pt,preprint]{aastex}
\usepackage{natbib}
\newcommand{\etal}{{et al.~}}
\newcommand{\mmo}{\ensuremath{ {(m-M)}_{0} }}
\newcommand{\ebv}{\ensuremath{E(B-V)} }
\def\lea{\mathrel{<\kern-1.0em\lower0.9ex\hbox{$\sim$}}}
\def\gea{\mathrel{>\kern-1.0em\lower0.9ex\hbox{$\sim$}}}

\begin{document}


\title{HST/ACS Photometry of Old Stars in NGC 1569: The Star Formation 
History of a Nearby Starburst\altaffilmark{1}}


\author{Aaron J. Grocholski\altaffilmark{2,3,4}, Roeland P. van der 
Marel\altaffilmark{2}, Alessandra Aloisi\altaffilmark{2}, Francesca 
Annibali\altaffilmark{5}, Laura Greggio\altaffilmark{6}, Monica 
Tosi\altaffilmark{5}}

\altaffiltext{1}{Based on observations with the NASA/ESA {\it Hubble
Space Telescope}, obtained at the Space Telescope Science Institute,
which is operated by AURA for NASA under contract NAS 5-26555}

\altaffiltext{2}{Space Telescope Science Institute, 3700 San Martin Dr.,
Baltimore, MD 21218, USA; aloisi, marel@stsci.edu}
 
\altaffiltext{3}{Astronomy Department, Yale University, New Haven, CT,
06520, USA}

\altaffiltext{4}{current address: Department of Astronomy, University of 
Florida, Gainesville, FL 32607, USA; a.grocholski@astro.ufl.edu}
 
\altaffiltext{5}{Osservatorio Astronomico di Bologna, INAF, Via Ranzani
1, I-40127 Bologna, Italy; francesca.annibali, monica.tosi@oabo.inaf.it}

\altaffiltext{6}{Osservatorio Astronomico di Padova, INAF, vicolo 
dell'Osservatorio 5, 35122 Padova, Italy; laura.greggio@oapd.inaf.it}

\begin{abstract}

We used HST/ACS to obtain deep $V$- and $I$-band images of NGC 1569, one 
of the closest and strongest starburst galaxies in the Universe.  These 
data allowed us to study the underlying {\it old} stellar population, 
aimed at understanding NGC 1569's evolution over a full Hubble time. We 
focus on the less-crowded outer region of the galaxy, for which the 
color-magnitude diagram (CMD) shows predominantly a red giant branch 
(RGB) that reaches down to the red clump/horizontal branch feature 
(RC/HB).  A simple stellar population (SSP) analysis gives clear 
evidence for a more complicated star formation history (SFH) in the 
outer region.  We derive the full SFH using a newly developed code, 
SFHMATRIX, which fits the CMD Hess diagram by solving a non-negative 
least squares problem.  Our analysis shows that the relative 
brightnesses of the RGB tip and RC/HB, along with the curvature and 
color of the RGB, provide enough information to ameliorate the 
age-metallicity-extinction degeneracy.  The distance/reddening 
combination that best fits the data is \ebv = 0.58 $\pm$ 0.03 and $D = 
3.06 \pm 0.18$ Mpc. Star formation began $\sim$ 13 Gyr ago, and this 
accounts for the majority of the mass in the outer region. However, the 
initial burst was followed by a relatively low, but constant, rate of 
star formation until $\sim$ 0.5-0.7 Gyr ago when there may have been a 
short, low intensity burst of star formation. Stellar metallicity 
increases over time, consistent with chemical evolution expectations. 
The dominant old population shows a considerable spread in metallicity, 
similar to the Milky Way halo.  However, the star formation in NGC 
1569's outer region lasted much longer than in the Milky Way. The 
distance and line-of-sight velocity of NGC 1569 indicate that it has 
moved through the IC 342 group of galaxies, which may have caused this 
extended star formation. Comparison with other recent work provides no 
evidence for radial population gradients in the old population of NGC 
1569, suggesting that our results are representative of the old stellar 
population throughout the galaxy.

\end{abstract}


\section{Introduction}
\label{intro}

At high redshift, the star formation rate density is considerably higher 
than what we observe in the local Universe (e.g., 
\citealt{hopkinsandbeacom2006}). Many galaxies, such as Lyman break 
galaxies, are thought to be undergoing massive bursts of star formation. 
These starbursts are important to understand since they drive the 
evolution of galaxies. Energy generated in a starburst provides thermal 
and mechanical heating to the host galaxy, and supernova winds spread 
chemically enriched material throughout the ISM.  In some cases, 
starbursts have enough energy to create a galactic wind that escapes the 
gravitational potential of the host galaxy and thereby disperses metals 
throughout the intergalactic medium (e.g., \citealt{veilleuxetal2005}).  

Given the unresolved nature of high redshift galaxies, the information 
we can glean from them has its limits.  One approach to understand 
the physical processes at work during the starburst phase of high 
redshift galaxies is to turn to studies of nearby starburst galaxies 
(although these may represent quite a different mass and evolutionary 
stage).  Although they are rare, the proximity of starburst galaxies in 
the local Universe allows us to resolve them into individual stars and 
thereby study their star formation histories (SFHs) in great detail over 
the lifetime of the galaxy.

A starburst is loosely defined as an intense period of star formation 
that is unsustainable over a Hubble time due to the limited supply of 
gas available in the galaxy.  Starbursts in the local Universe are 
typically found in dwarf irregular (dIrr) and blue compact dwarf (BCD) 
galaxies, systems which are characterized by their relatively blue 
colors due to young stellar populations, high gas content, and low 
metallicities.  In some cases, their metal abundances are so low that 
these galaxies have been targeted as possible primeval galaxies that may 
have only recently begun forming stars \citep{izotovandthuan1999}.  
However, in all dIrrs and BCDs that have been observed to a sufficient 
photometric depth, red giant branch stars (RGB; i.e., evolved stars 
older than $\sim$ 1 Gyr) have been detected, proving that these galaxies 
have been forming stars for an extended period of time (e.g., I Zw 18, 
\citealt{aloisietal2007}; SBS 1415+437, \citealt{aloisietal2005}; I Zw 
36, \citealt{schulteladbecketal2001}).

The dIrr galaxy NGC 1569 is one of the closest and most extreme 
starbursts.  NGC 1569 has a total dynamical mass of $M = 3.3 \times 
10^8$ M$_{\odot}$, 1/3 of which is in H{\small I} \citep{israel1988}. 
With a mean oxygen abundance of 12 + log($O/H$) $= 8.2 \pm 0.2$ (or $Z = 
0.25$Z$_{\odot}$, assuming [O/Fe] = 0.0; \citealt{greggioetal1998} and 
references therein), NGC 1569 is thus a gas-rich system with an SMC-like 
chemical composition.  Long assumed to lie at a distance of 2.2 Mpc 
\citep{israel1988}, in \citet{grocholskietal2008} we unequivocally 
identified for the first time the tip of the red giant branch (TRGB) and 
found that its true distance is $\sim$ 3 Mpc (see also \S 5.1).  This 
makes it a likely member of the IC 342 group of galaxies.  In the past 
$\sim$ 25 Myr, NGC 1569 has formed a large number ($\sim$ 150) of star 
clusters with masses similar to Milky Way open clusters 
(\citealt{andersetal2004}; see also \citealt{origliaetal2001}, 
\citealt{hunteretal2000}), as well as three super star clusters.  Based 
on our improved distance, the three super star clusters, NGC 1569-A1 and 
NGC 1569-A2 (\citealt{deMarchietal1997}, \citealt{gilbertgraham2002}) 
and NGC 1569-B (\citealt{larsenetal2008}) have masses of (6-7) $\times 
10^5$ M$_{\odot}$.  These clusters, which are more massive than any of 
the young clusters in the Milky Way or Large Magellanic Cloud, will 
likely survive for the entire lifetime of the galaxy, thus making them 
possible precursors to globular clusters.  Many authors have studied the 
recent star formation in NGC 1569 using various tracers of the ISM, 
including: CO (\citealt{youngetal1984}; \citealt{greveetal1996}; 
\citealt{tayloretal1999}); HI (\citealt{israelandvandriel1990}; 
\citealt{stilandisrael1998}); H$\alpha$ (\citealt{hunteretal1993}; 
\citealt{devostetal1997}); and the X-ray emission from hot gas 
(\citealt{heckmanetal1995}; \citealt{dellacecaetal1996}; 
\citealt{martinetal2002}).  All of these indicators point towards an ISM 
heavily impacted by a starburst driven galactic superwind.

Star formation in the central region of NGC 1569 has been studied by a 
number of authors using HST photometry (e.g., 
\citealt{vallenariandbomans1996, greggioetal1998, angerettietal2005}).  
The most straightforward way to determine a galaxy's SFH is by comparing 
its observed color-magnitude diagram (CMD) to synthetic CMDs, created 
from stellar evolution models, with star formation rates that vary over 
time.  Both \citet{greggioetal1998} and \citet{angerettietal2005} used 
this method to determine the SFH of NGC 1569 and found that in the past 
$\sim$1 Gyr, NGC 1569's star formation rate per unit area is 2-3 times 
higher than in other strong starbursts and 2-3 orders of magnitude 
higher than what is seen in Local Group irregulars and the solar 
neighborhood.  At such a rate, the star formation in NGC 1569 would have 
exhausted its gas supply in $\sim$1 Gyr.  Thus, to have sustained star 
formation for an extended period of time, NGC 1569 would need to have 
accreted gas, possibly from a nearby HI cloud 
(\citealt{muhleetal2005,stilandisrael1998}).  However, since neither 
program's data reached the faint magnitudes needed to sample the RGB, 
they were not able to constrain NGC 1569's SFH prior to 1 Gyr ago.

While the existence of an RGB indicates stars older than 1 Gyr, the 
age-metallicity degeneracy of these stars makes it difficult to resolve 
the age distribution of stars from the RGB alone. The most accurate way 
to determine the SFH of any stellar population is by resolving the main 
sequence turn off (MSTO) for even the oldest stars. However, at $M_I 
\sim 4$ for 13 Gyr old stars, the MSTO is too faint to be observed in a 
reasonable amount of telescope time for galaxies outside of about 1 Mpc. 
So instead, we must turn to brighter CMD features, such as the core 
helium burning stars of the red clump/horizontal branch (RC/HB; $M_I 
\sim -0.2$) to help lift the age-metallicity degeneracy.   
\citet{rejkubaetal2005} and \citet{rejkubaetal2011} analyzed a CMD of 
the elliptical galaxy Cen A (NGC 5128) that reached 0.5 mag below the 
RC/HB and showed that photometry reaching these depths can provide 
enough information to determine a galaxy's SFH over a Hubble time. 
Therefore, to determine the SFH of NGC 1569's oldest stars, we obtained 
deep HST ACS/WFC $V$- and $I$-band images that reach down to the RC/HB 
and cover both the crowded inner and relatively sparsely populated outer 
regions (HST GO-10885, PI: A. Aloisi).  Our observational program 
was designed to reach $\sim$ 0.5 mag below the RC/HB.  However, since 
NGC 1569 is roughly 50\% farther away than previously believed, our 
photometry only barely reaches the RC/HB and not well below it. 
Recently, as a part of a study of 18 nearby dwarf starburst galaxies, 
\citet{mcquinnetal2010} analyzed a subset of our images.  Using the CMD 
fitting program, MATCH (\citealt{dolphin2002}), they determined the full 
SFH of NGC 1569.  Although they focused primarily on the starburst 
properties, they also determine a relatively coarse age distribution for 
stars older than 1 Gyr.  We will discuss their results further in \S 
5.3.

\begin{figure}
\plotone{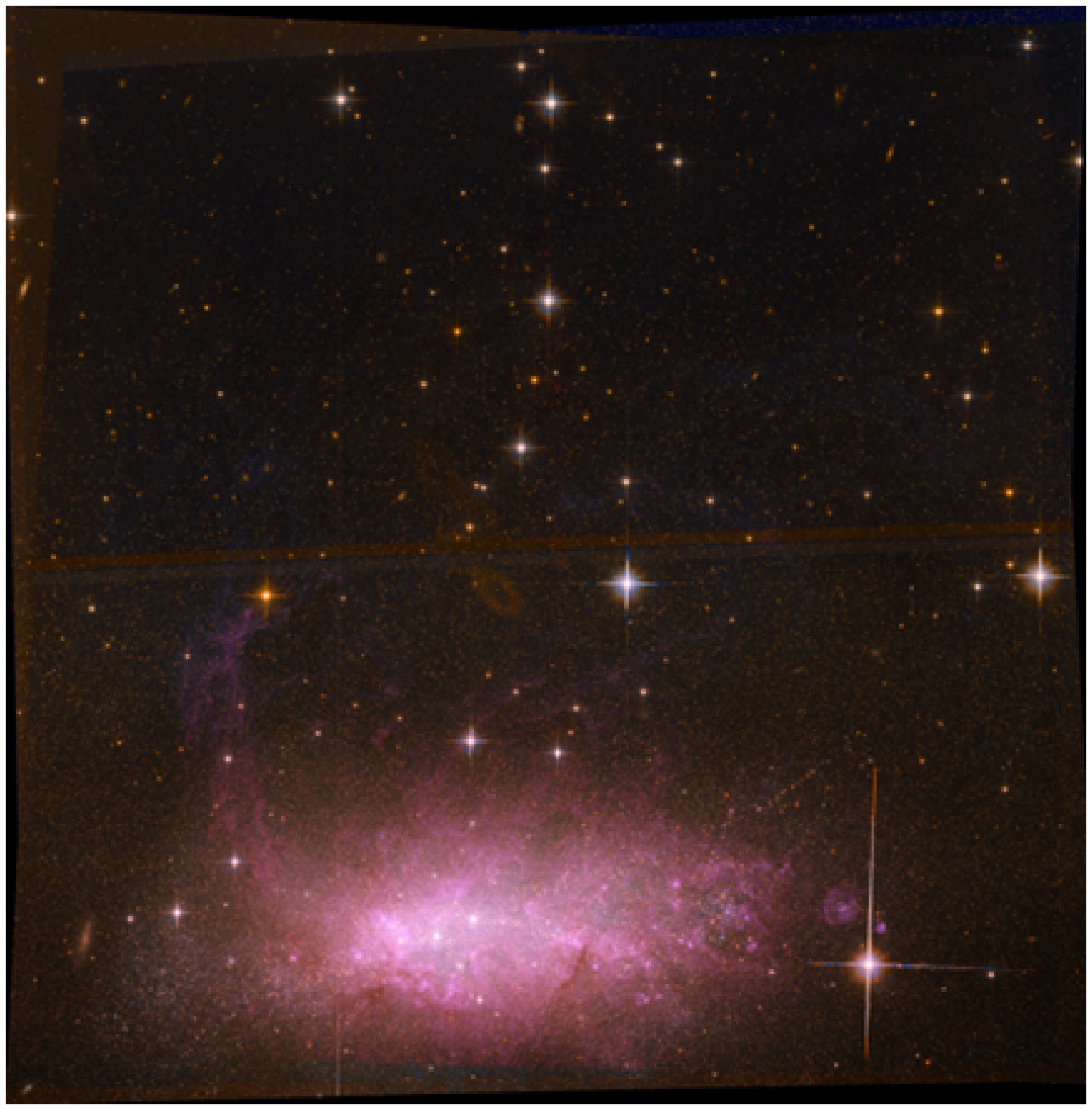}
\caption{ACS/WFC 3-color image of NGC 1569, where the H$\alpha$-, $I$-
and $V$-bands have been colored red, orange, and blue, respectively.
(The faint ellipse near the center of the field is the ghost image of a
bright star).}
\label{ngc1569_color}
\end{figure}

Since the SFH of young stars in NGC 1569 has been well studied by 
previous authors, in this paper we present an analysis focused on the 
SFH of stars older than 1 Gyr.  In \citet{grocholskietal2008} we showed 
that the recent star formation is concentrated in the core of NGC 1569, 
while the outer region shows no signs of a significant young population.  
In \citet{grocholskietal2008} we adopted the convention of 
referring to the bottom half of our NGC 1569 field 
(Fig.~\ref{ngc1569_color}) as the core and the top half as the halo.  
Recently, in \citet{rysetal2011}, we used four WFPC2 fields to study the 
number density profile of RGB stars in NGC 1569 out to 8 scale radii and 
found that NGC 1569 does appear to transition from an exponential disk 
to a halo population.  However, as this occurs outside of our ACS field, 
herein we refer to the upper half of our ACS field as the outer region 
of NGC 1569.  Due to the fact that the outer region is much less 
crowded than the core, these data necessarily go deeper, allowing us to 
more accurately detect and measure the brightness of stars on the RC/HB.  
This is illustrated by the luminosity functions (LFs) in Fig.~4 of 
\citet{grocholskietal2008}, where the peak due to the RC/HB is visible 
in the outer region, but not in the core. Therefore, herein we focus 
solely on our observations of NGC 1569's outer region.

In \S 2 we present our data and detail the process used to determine the 
photometry and errors.  The resulting CMDs for the core and outer 
region are discussed in \S 3.  In \S 4 we test whether or not NGC 
1569's outer region can be treated as a simple stellar population 
(i.e., if it can be represented by a single age, single metallicity 
population) while in \S 5 we use the method of synthetic CMD fitting to 
determine the full SFH of NGC 1569 and compare our results to the work 
of \citet{mcquinnetal2010}. In \S 6 we discuss NGC 1569's SFH in the 
context of interactions with the IC 342 group of galaxies. Conclusions 
are summarized in \S 7. Appendix~A provides an overview of our new code 
SFHMATRIX for determining the SFH a galaxy.

\section{Imaging Data}
\label{data}
\subsection{HST/ACS Observations}
\label{obs}

We observed NGC 1569 with the HST ACS/WFC in November 2006 and January
2007 as a part of HST program GO-10885 (PI: Aloisi).  Images were taken
in the F606W ($V$) and F814W ($I$) broad-band filters as well as the
F658N (H$\alpha$) narrow-band filter.  Total exposure times were: 61716
s in the $V$-band, composed of 54 individual images; 24088 s in the
$I$-band, composed of 22 images, and 4620 s in the H$\alpha$ filter,
composed of 4 images.  While all of the images were centered on the
galaxy, telescope rotation between the two observation dates caused both
the $V$- and $I$-band images to be split between two different
orientation angles.  NGC 1569 was only imaged in H$\alpha$ during the
second visit and, therefore, these data are only at one orientation. 
WFPC2 was used in parallel to image regions in the outskirts of NGC
1569, $\sim6\arcmin$ from the galaxy's center, also in the $V$- and
$I$-bands.  These data are discussed in \citet{rysetal2011}.

Our ACS images were dithered using a standard sub-pixel plus integer
dither pattern.  The sub-pixel pattern is used to improve the sampling
of the point spread function (PSF) and aid in the removal of bad/hot
pixels and cosmic rays.  The integer pixel step is necessary to fill in
the gap between the ACS chips.  While these are the same data that were
presented in \citet{grocholskietal2008}, we have reprocessed all of the
images using the most up-to-date versions of the ACS pipeline (CALACS)
and calibration frames.  Images in each filter were then combined into a
single image using the MULTIDRIZZLE software package
(\citealt{fruchteretal2009}).  The MULTIDRIZZLE software fine-tunes the
image alignment, corrects for small shifts, rotations, and geometric
distortions between the images, and removes cosmic rays and bad pixels. 
We experimented with the drizzle parameters within MULTIDRIZZLE and
found that resampling the images to 0.7 times the original ACS/WFC pixel
scale provides the best resolution and PSF sampling for our data.  Our
final combined images have a pixel size of $0.035\arcsec$ and cover
roughly $3.5\arcmin \times 3.5\arcmin$.  Figure \ref{ngc1569_color}
shows our 3-color image of NGC 1569.  The H$\alpha$ image was not used
for photometry and is thus excluded from further discussion in this 
paper.

\subsection{Photometry and Calibration}

Using the stand-alone versions of DAOPHOT and ALLSTAR
(\citealt{stetson1987}), we performed photometry on the $V$- and
$I$-band images in the following manner.  We created a rough PSF model
from $\sim$300 bright, uncrowded stars in each image.  This rough model
was then used to remove neighbors from around the full set of $\sim$1000
PSF stars in each image, thereby allowing us to create a more robust
PSF.  Our PSF stars were chosen to have excellent spatial coverage
across the entire image so as to accurately model variations in the PSF
shape as a function of position.  We note that, due to crowding effects,
we have avoided using any stars in the crowded inner region of NGC 1569. 
Next, we used ALLSTAR to fit the improved PSF model to independent
source detection lists for the $V$- and $I$-band images.  In an effort
to find and measure faint stars, we performed a single iteration of
subtracting from the images all sources measured in the first ALLSTAR
run, searching for previously undetected objects, adding those objects
to the original source detection list, and then re-running the PSF
fitting on the original images, this time using the newly updated
catalog. 

We matched the $V$ and $I$ photometry lists, requiring that detections
were within 0.5 pixel to be considered a match.  The resulting
photometric catalog contains over 400,000 sources.  To further clean our
catalog of false detections or background galaxies, we made cuts by both
position and photometric quality.  Positional cuts were used to trim
detections from around overexposed stars, where diffraction spikes can
lead to false detections, as well as from the edges of the images and
the ACS chip gap, which had lower exposure time due to dithering and
therefore a much higher incidence of noise peaks that were coincident
between the $V$ and $I$ images.  ALLSTAR provides three estimates of the
quality of the PSF fit to any given source: $\sigma$, the error in the
PSF magnitude; $\chi$, the goodness-of-fit for the PSF fit to each star;
and $sharpness$, a measure of the intrinsic size of the source relative
to the PSF.  Assuming that noise peaks follow a Gaussian distribution
around the mean background level, we expect on the order of one noise
peak that meets our matching criteria and lies 3.5$\sigma$ above the
mean in {\it both} bands.  Since a signal-to-noise ratio of 3.5
translates to an error of $\sim$0.3 mag, we keep only those stars with
$\sigma_{V} \le 0.3$ {\it and} $\sigma_{I} \le 0.3$.  Sources with a
$sharpness$ $\ll$ 0 are considerably more narrow than the PSF and are
likely to be false detections, while those with $sharpness$ $\gg$ 0 are
more extended and likely to be background galaxies or stellar blends. 
Thus, we cut from our catalog any stars with $|sharpness| > 1$ in
either band.  We have chosen to not make any cuts based on $\chi$ since
most of the faint objects with large $\chi$ values have already been cut
based on their $\sigma$ and $sharpness$ values.  Bright sources with 
large $\chi$ values are generally real stars, with the large $\chi$ due 
to inadequacies in the PSF (which appear more significant when the star 
is bright and the noise is low). After applying all of the above cuts, 
our photometric catalog contains over 370,000 stars, with 
approximately 31,000 of those stars residing in the outer region.

Finally, we apply the necessary zeropoints and corrections to place our
photometry on the Johnson-Cousins magnitude system following the
prescriptions in \citet{siriannietal2005}.  First, we converted
instrumental magnitudes to the HST VEGAMAG system according to the
equation
\begin{equation}
m_{vegamag} = m_i + C_{ap} + ZP - C_{inf} + C_{sky} + C_{cte}.
\end{equation}
$m_i$ is the PSF-fitting magnitude [$-2.5 \times log(flux)$] within 
a radius of 10 pixels (0.35\arcsec) and $C_{ap}$ is the correction
necessary to convert that magnitude to the conventional 0.5\arcsec
~radius aperture.  ZP is the HST VEGAMAG
zeropoint\footnote{http://www.stsci.edu/hst/acs/analysis/zeropoints} for
a given filter and $C_{inf}$ is an offset to convert magnitudes from the
0.5\arcsec~radius aperture to an infinite aperture 
\citep{siriannietal2005}.  $C_{sky}$ corrects for the fact that our 
aperture correction, $C_{ap}$, uses an annulus that is a finite 
distance from the star to estimate the sky background.  Since a small 
amount of star light is included in the annulus, the sky background 
is overestimated in the aperture correction step.  The ACS CCDs suffer 
from losses in charge transfer efficiency (CTE) as a result of their 
prolonged exposure to radiation in space.  We corrected for CTE losses 
following the prescription in \citet{riessandmack2004}, namely, 
\begin{equation}
C_{cte} = 10^A \times SKY^B \times FLUX^C \times \frac{Y}{2048} \times 
\frac{MJD - 52333}{365},
\end{equation}
where $SKY$ is the sky background counts per pixel, $FLUX$ is the star
counts per exposure within our PSF fitting radius, $Y$ is the number of
charge transfers, and $MJD$ is the Modified Julian Date.  The
coefficients were taken from \citet{riessandmack2004} for an aperture
radius of $r = 7$ pixels on the native pixel scale (or 10 pixels on our
resampled drizzled image), and are $A = -0.7$, $B = -0.34$, and $C =
-0.36$.  Due to the fact that our images were both dithered and taken at
two different orientations, each star has 54 different charge transfer
values, $Y$, in the $V$-band and 22 different $Y$ values in the
$I$-band.  We tested two approaches for dealing with this in calculating
$C_{cte}$ for each star.  First, we calculated the CTE corrections for
each star in each individual image and averaged those together to get
$C_{cte}$ for each star.  In the second approach, we determine the
average number of charge transfers for each star and then calculate a
single CTE correction for each star.  We found that these two approaches
give similar $C_{cte}$ values for each star.  Our average CTE correction
is $\sim$ 0.01 mag, but can be as large as $\sim$ 0.2 mag for the
faintest stars.  After applying all of the zeropoints and corrections,
the last step in our calibration was to convert photometry on the
VEGAMAG system to the Johnson-Cousins system by following the procedure
outlined in \citet{siriannietal2005}.  Note that we have not applied any
reddening corrections to our photometry since these are included as part 
of our models (see \S\ref{LFs} and~\S\ref{sfh}).

\subsection{Completeness and Errors}

To properly evaluate the role of incompleteness and photometric errors 
in the analysis of our data, we performed artificial star tests on our 
$V$- and $I$-band images.  We added artificial stars to our images using 
the ADDSTAR routine within DAOPHOT.  This routine simulates real stars 
covering a user supplied range of positions and magnitudes by adding the 
appropriate Poisson noise to the previously generated PSFs.  In an 
effort to compromise between efficient computing of the completeness and 
not changing the crowding on the images, we divided the images into a 
grid of boxes 30 pixels $\times$ 30 pixels in size and placed one 
artificial star in each box, with the added restriction that no two 
artificial stars lie within two PSF fitting radii (20 pixels) of each 
other.  To fully sample the images, we varied both the starting position 
of the grid as well as the position of the artificial star within each 
box, allowing for shifts as small as 0.01 pixels.  We chose the range of 
magnitudes covered by the artificial stars to extend $\sim$ 1 mag 
brighter and fainter than the observed data.  To better mimic the 
observed luminosity function, we generated twice as many artificial 
stars in the faint half of the magnitude range as in the bright half. 
After placing the artificial stars on the image, we followed the same 
procedure as outlined in the previous section and performed PSF fitting 
on the entire image.  We then cross-correlated the input artificial star 
list with the output photometric catalog, again using a 0.5 pixel 
matching radius.  An artificial star was considered to be ``lost" if it 
was not matched in the output catalog, or if its PSF fitting magnitude 
differs from its input magnitude by more than 0.75 mag, it has $\sigma > 
0.3$ or $|sharpness| > 1$.  We note that we also trimmed the artificial 
stars by position in the same way as for the observed catalog, but these 
stars were not counted in the lost/recovered statistics.  In the outer 
region, we simulated over 170,000 stars in each band, or more than 5 
times the number of stars observed in the outer region.

\section{Color-Magnitude Diagrams}
\label{cmd}

\begin{figure}
\plotone{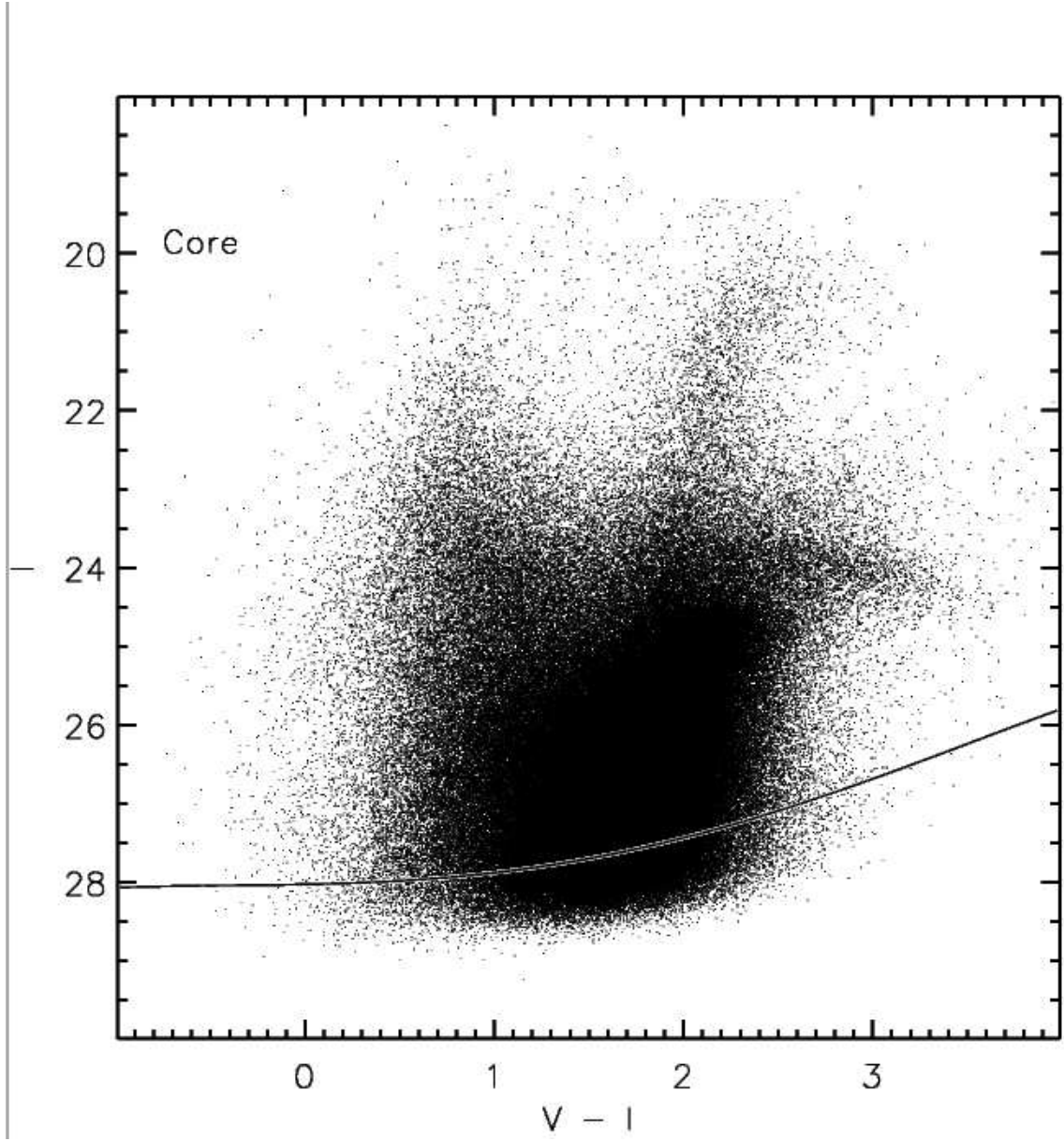}
\caption{CMD composed of all stars in the core of NGC 1569 (bottom half 
of Fig.~\ref{ngc1569_color}).  Features 
resulting from young, massive, main sequence and evolved stars dominate 
the CMD (see \S3 for details).  The 20\% completeness limit is shown as
the solid line.  The remainder of this paper focuses 
solely on the outer region of NGC 1569.
}
\label{core_cmd}
\end{figure}

\begin{figure}
\plotone{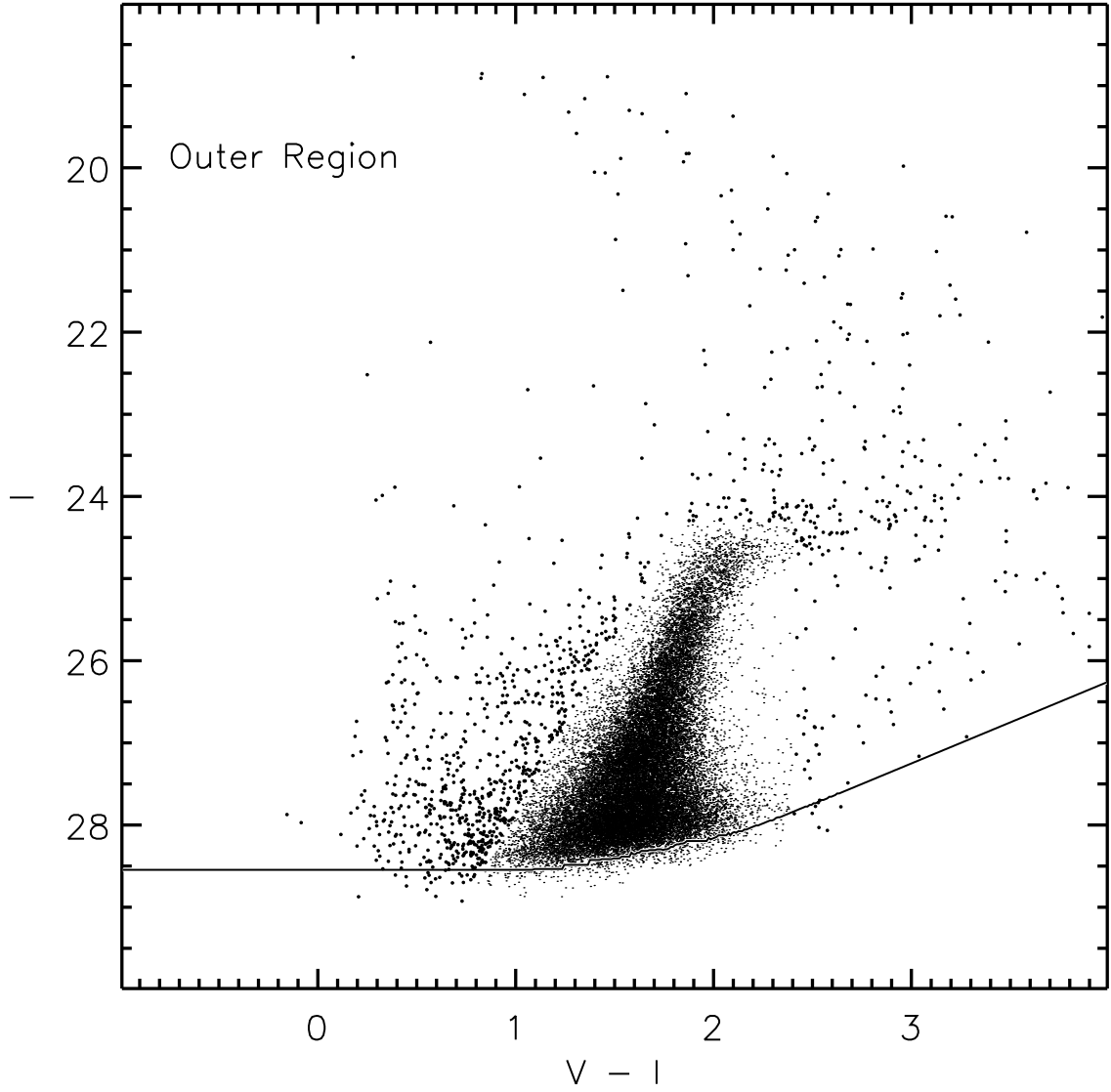}
\caption{CMD of all stars in NGC 1569's outer region (top half 
of Fig.~\ref{ngc1569_color}).  Unlike the core CMD, 
only the old stars on the RGB are readily visible in the outer region, 
illustrating that the recent star formation in NGC 1569 is concentrated 
in the core.  For visibility, we have increased the size of the points
for stars that are off of the RGB.  As in the previous figure, the
solid line represents the 20\% completeness limit.
}
\label{halo_cmd}
\end{figure}

In Figs.~\ref{core_cmd} and \ref{halo_cmd} we present the resulting CMDs 
for the core and outer region of NGC 1569. Previous authors have shown 
that NGC 1569 has recently undergone massive bursts of star formation 
(e.g., \citealt{greggioetal1998,angerettietal2005}).  This is evident in 
the CMD of the core (Fig.~\ref{core_cmd}), which is dominated by 
features that are due to young stellar populations.  Readily visible 
above $I \sim 24$ and with $0.6 \la (V-I) \la 1.1$ is the blue plume, 
which contains both young ($\la$ 10 Myr) main sequence stars and massive 
evolved stars ($\ga$ 9 M$_{\odot}$) on the blue part of their core 
helium-burning phase.  The red plume of supergiants at $1.9 \la (V-I) 
\la 2.5$ and the blue loop stars (between the red and blue plumes) 
indicate the existence of evolved stars with masses $\ga$ 5 M$_{\odot}$.  
Also visible at $I \sim 24$ and $(V-I) \ga 2.5$ are the 
intermediate-mass ($\sim$ 1.2 - 6 M$_{\odot}$) carbon stars of the 
thermally pulsating asymptotic giant branch (AGB), as well as M-type AGB 
stars.

Unlike the core, the CMD of the outer region (Fig.~\ref{halo_cmd}) shows 
no signs of a significant young stellar population, indicating that the 
recent bursts of star formation in NGC 1569 were restricted to the core.  
In the outer region, the only outstanding feature is the upper $\sim$ 4 
mag of the red giant branch (RGB), which is the result of stars older 
than $\sim 1$ Gyr that have evolved off of the main sequence and are 
expanding as their He cores increase in mass, contract, and work their 
way toward becoming fully degenerate.  RGB stars are also present in the 
core, though heavily blended in the CMD with younger evolutionary 
features, indicating a global star formation history that began at least 
1 Gyr ago.

\begin{figure}
\plotone{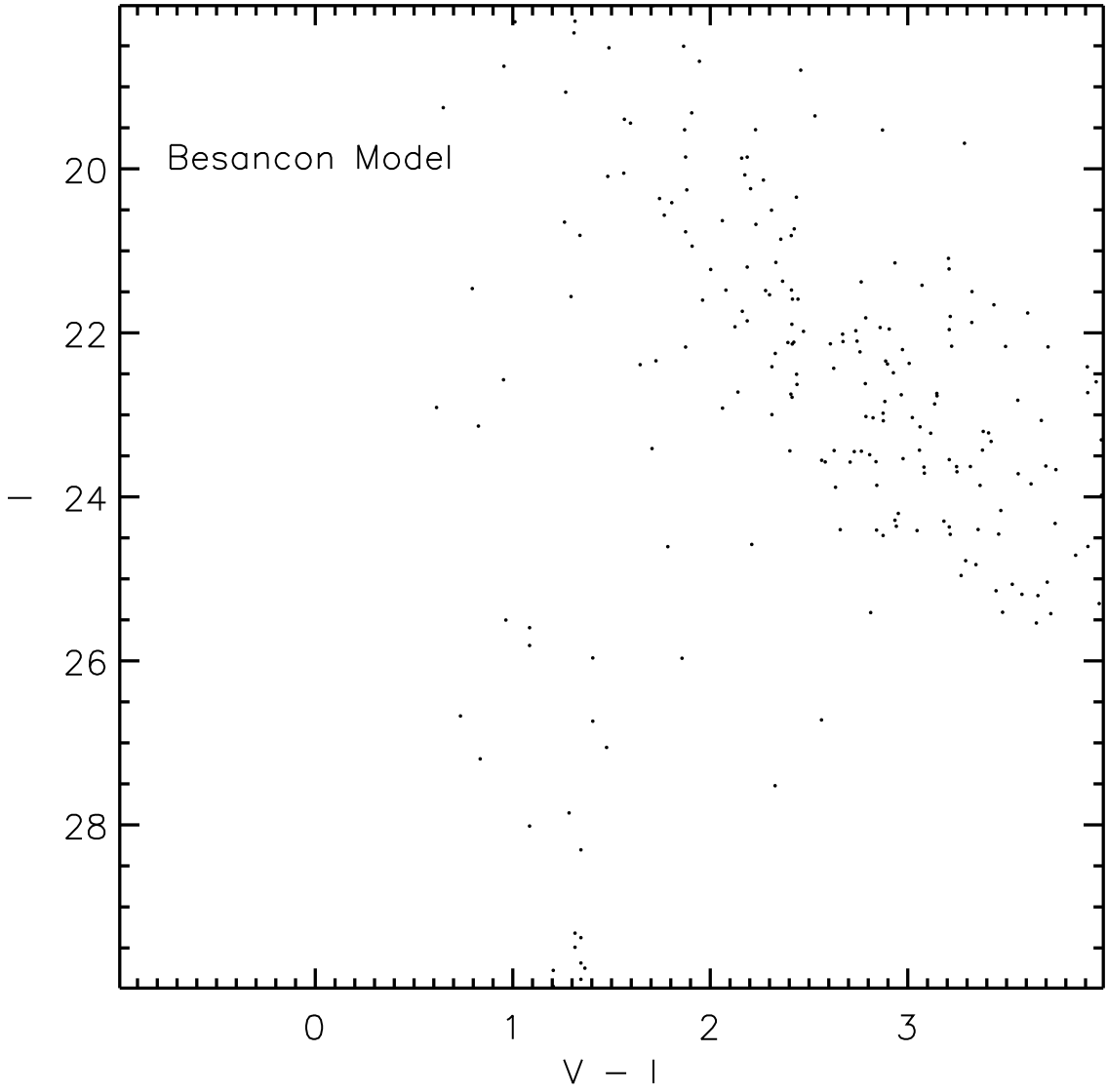}
\caption{Besan\c{c}on model of the expected MW foreground contamination 
in the direction of NGC 1569 \citep{besancon}, for a field size equal 
to that of the outer region (i.e., half of the total HST/ACS field).
}
\label{besancon}
\end{figure}

Given NGC 1569's location ($\ell = 143\fdg68213$, $b = 11\fdg24174$) 
near the Galactic Plane, we expect our CMDs to suffer from some 
foreground contamination due to the Galaxy.  This contamination can be 
seen in Fig.~\ref{halo_cmd} as the swath of stars stretching from $I 
\sim 19$, $(V-I) \sim 1$ to $I \sim 25$, $(V-I) \sim 4$.  In 
Fig.~\ref{besancon}, we confirm that this ``feature" is due to the Milky 
Way by plotting a CMD of the expected foreground stars in our ACS field 
based on the Besan\c{c}on model of the Galaxy \citep{besancon}.  The 
Besan\c{c}on model includes four populations: the thin disk, thick disk, 
bulge, and spheroid.  The plume of faint stars around $V-I \sim$ 1.3 is 
due to white dwarfs in the disk, and the stars in top right part of the 
CMD are primarily M-dwarfs in the disk.  These foreground stars are not 
considered further in the remainder of this paper.  We note that after 
taking into account the foreground stars, there is a sparse blue ($V-I 
\sim 0.4$ in Fig.~\ref{halo_cmd}) plume of stars in the outer region 
that are not related to the old RGB stars.  These stars are spread 
throughout the outer region, but with a higher concentration toward the 
core of the galaxy, and are possibly young stars that have migrated out 
of the core (\S 5.3).

\section{Luminosity Function Analysis with Simple Stellar Population Models}
\label{LFs}

As we have shown above, the outer region of NGC 1569 appears to be a 
purely
old population and thus provides the opportunity to study the SFH of
stars older than $\sim$ 1 Gyr.  Since we have no {\it a priori} reason
to assume a specific SFH for the outer region (e.g., that it formed in a
single burst or experienced extended episodes of star formation), we
begin with the most straightforward approach of treating NGC 1569's
outer region as a simple stellar population (SSP), i.e., one that can be 
fit
by a theoretical isochrone with a single age and metallicity.  Models
with more complicated SFHs are addressed in \S\ref{sfh}.  For both the
SSP analysis and full SFH analysis, we use stellar evolution models
from the Padova group\footnote{http://stev.oapd.inaf.it/cgi-bin/cmd}
(e.g., \citealt{marigoetal2008}) as they provide comprehensive
coverage of the stages of stellar evolution and are readily available
for a large range of ages, metallicities, and photometric systems.

The metallicity of an isochrone is $Z$. Throughout this paper we will 
often use instead the approximation $log(Z/Z_{\odot}) =$ [Fe/H], where 
the equality holds for a solar composition mix.  Like the Padova 
models, we adopt $Z_{\odot} = 0.019$ from work by 
\citet{andersandgrevesse1989}\footnote{There are many indications for a 
lower value of $Z_{\odot} = 0.0134$ as advocated by 
\citet{asplundetal2009}, but it has so far not been possible to bring 
such a lower number in agreement with Solar sound speed measurements 
from helioseismology.  Recent 3D modeling of the solar photosphere 
from \citet{caffauetal2011} gives $Z_{\odot} = 0.0153$, which moves in 
the direction of reconciling solar photosphere and helioseismology 
measurements.}

\subsection{Theoretical RGB Luminosity Function}

The RGB contains a plethora of information, with age, metallicity,
distance, and reddening all playing a role in the luminosity, shape and
color of the RGB.  For a large range of ages and abundances, the
$I$-band luminosity of the tip of the RGB (TRGB) is approximately
constant at $M_I \sim -4.0$ \citep[e.g.,][]{barkeretal2004}, thus acting
as a standard candle.  Unfortunately, while the shape of the RGB can
place some constraints on [Fe/H], the RGB color suffers from a well
known degeneracy, where different combinations of age, metallicity, and
reddening can create RGBs that are similar in appearance. 

\begin{figure}
\plotone{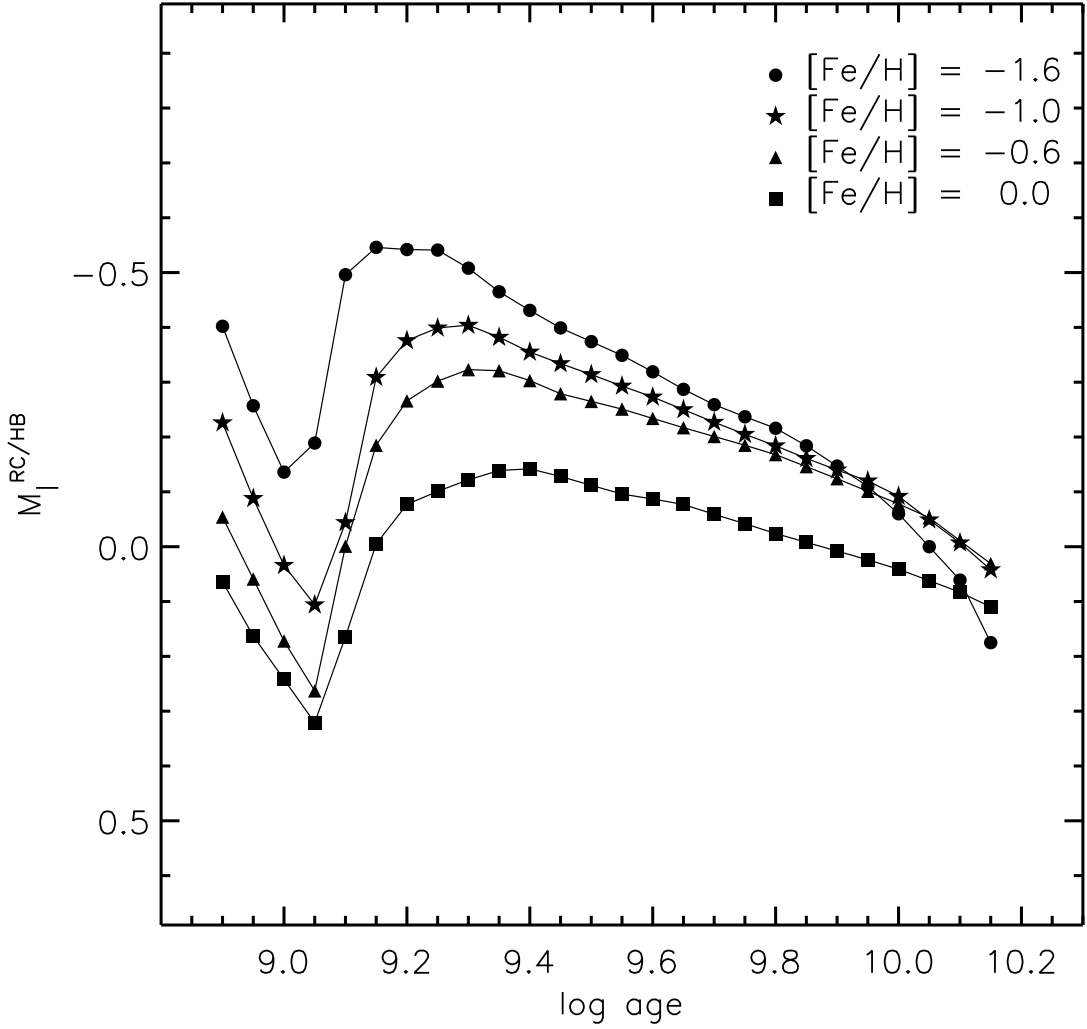}
\caption{Absolute $I$-band magnitude of the RC/HB feature plotted as a 
function of age for a range of metallicities.  For 
Figs.~\ref{rchb}-\ref{agbbump} we have used the theoretical values from 
the Padova isochrones.
}                
\label{rchb}
\end{figure}

Rather than dealing with the RGB as a whole, we instead turn to three 
features that may be visible in the LF of the RGB: the HB/RC, RGB bump, 
and AGB bump. The HB/RC feature on the CMD is a collection of evolved 
stars that are in their core helium burning phase. The exact luminosity 
of the RC/HB is dependent on both the age and metallicity of the stars 
(e.g., \citealt{gs02}, \citealt{girardisalaris2001}), but is roughly 
$3.5-4$ mag fainter than the TRGB in the $I-$band (Fig.~\ref{rchb}).

\begin{figure}
\plotone{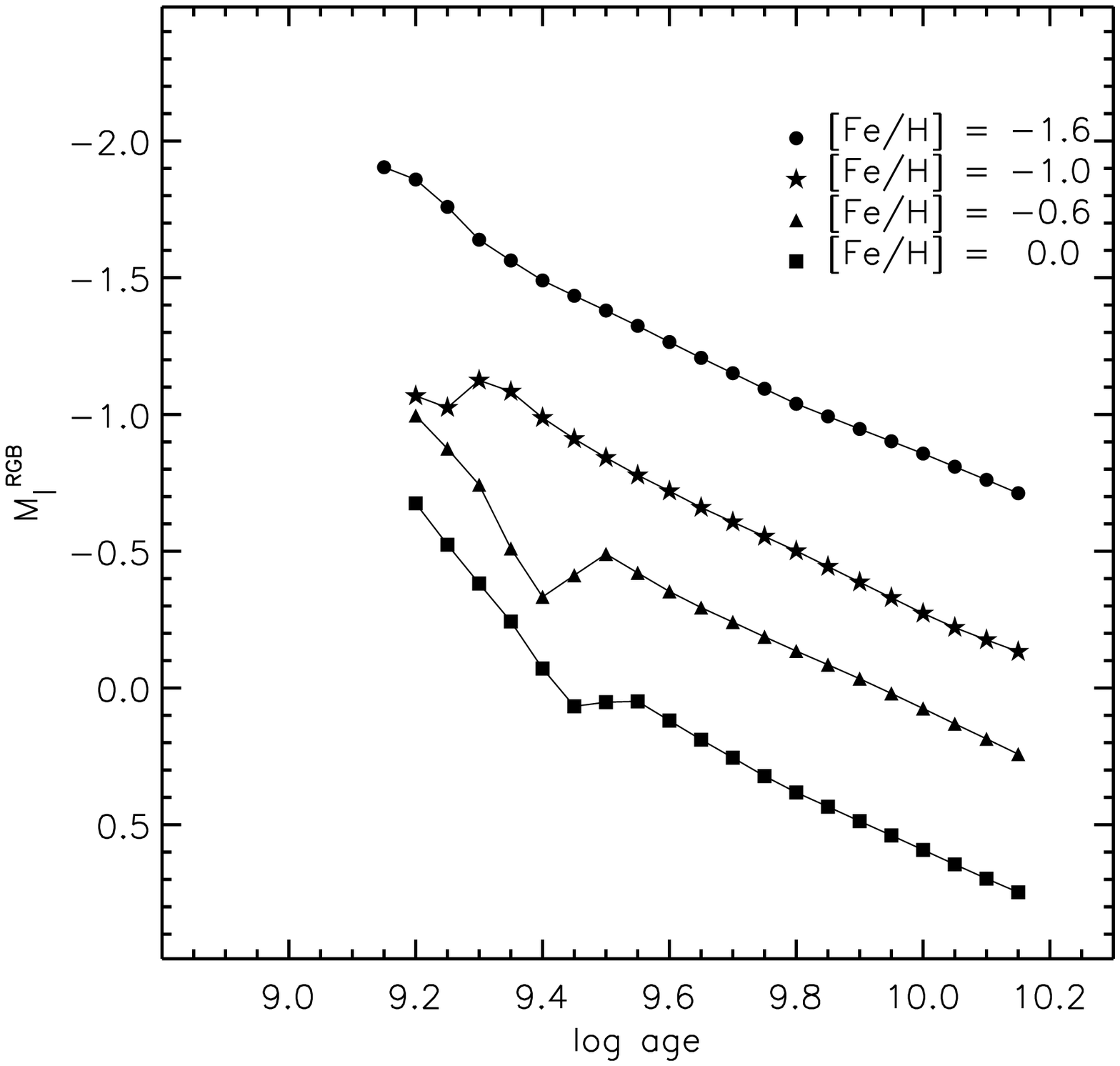}
\caption{Absolute $I$-band magnitude of the RGB bump plotted as a 
function of age for a range of metallicities.
}
\label{rgbbump}
\end{figure}

\begin{figure}
\plotone{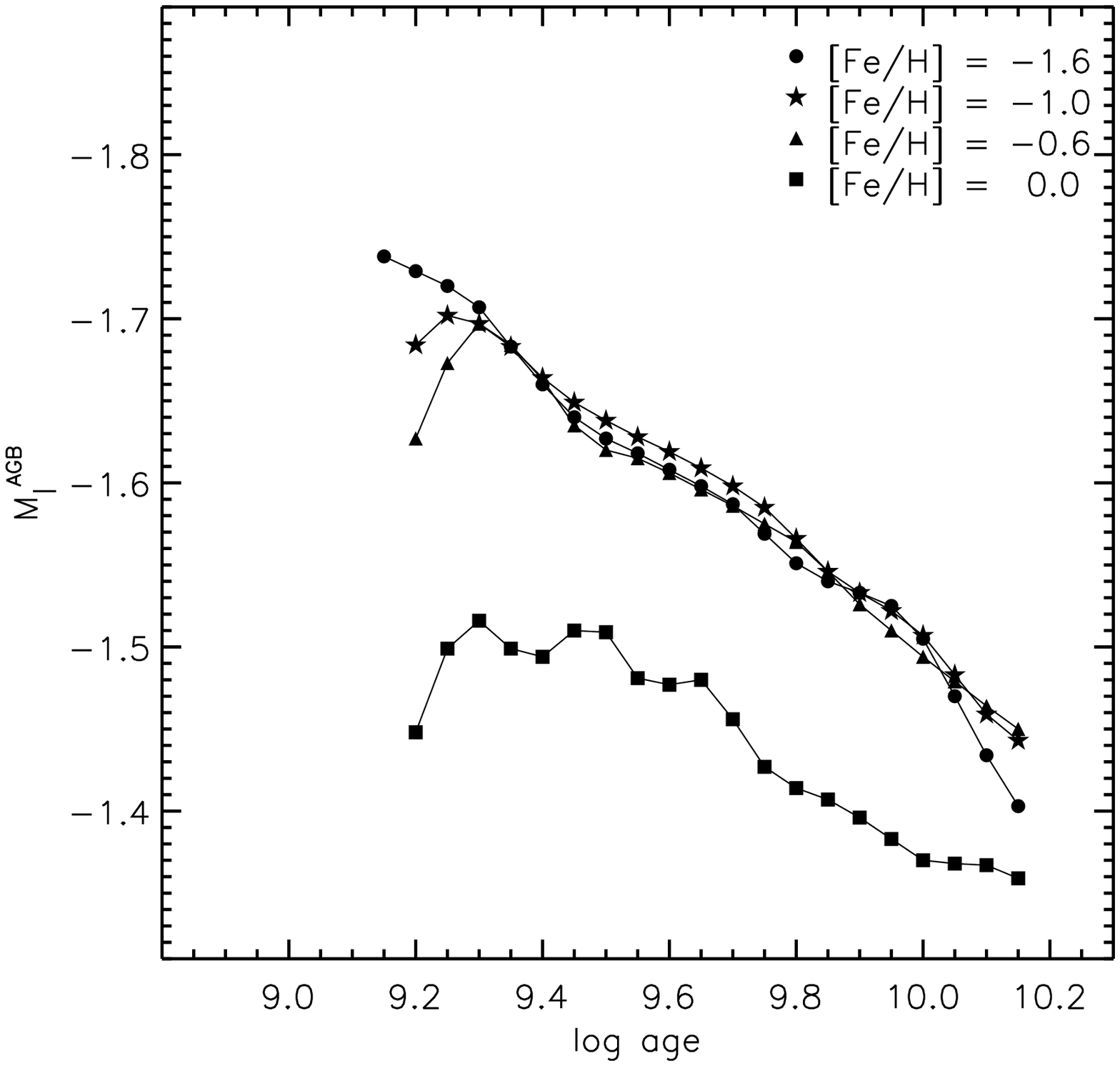}
\caption{Absolute $I$-band magnitude of the AGB bump plotted as a 
function of age for a range of metallicities.
}                             
\label{agbbump}
\end{figure}

Both the RGB bump and AGB bump are more subtle features that are 
typically not identifiable in the CMD of a stellar population and are 
only sometimes visible as discrete ``bumps" in the LF.  The RGB bump 
feature occurs when the H burning shell in RGB stars moves outward and 
crosses the chemical discontinuity produced during the first dredge up. 
Since the luminosity of the H burning shell, $L_H$, is related to 
the mean molecular weight, $\mu$, as $L_H \propto \mu^7$, the sudden 
drop in $\mu$ at the discontinuity (due to an increase in the H 
abundance) causes a temporary drop in the luminosity of an RGB star.  
RGB stars spend $\sim$20\% of their total RGB lifetime in the bump 
phase.  Due to the fact that the depth of the chemical discontinuity is 
dependent on both the mass and composition of the star, the luminosity 
of the RGB bump acts as a tracer of the age and metallicity 
(Fig.~\ref{rgbbump}).  Following central He exhaustion, RC/HB stars 
evolove up the AGB.  The transition from core helium burning to a thick 
He shell burning configuration causes a temporary drop in the luminosity 
of AGB stars, which results in the formation of the AGB bump.  
Fig.~\ref{agbbump} shows that the luminosity of the AGB bump is 
primarily dependent on the age of the population, with [Fe/H] playing a 
role only for the most metal rich systems.

Based on theoretical models, the RC/HB is visible in all stellar 
populations with log(age) $\ga$ 8.8 (630 Myr), regardless of [Fe/H].  On 
the other hand, while both the RGB bump and AGB bump features form in 
stellar populations with log(age) $\ga$ 9.1 (1.25 Gyr), these features 
cannot always be identified in a LF.  Aside from situations where the 
number of stars is so low that few stars exist on the RGB (e.g., open 
clusters), the significant range in brightness of the RGB bump ($>$ 2.5 
mag) can cause it to overlap with either the AGB bump or the RC/HB.  In 
contrast, the AGB bump varies in brightness by less than 0.5 mag and is 
never less than 1.0 mag brighter than the RC/HB.  This gives rise to 
stellar populations that can have either two or three distinct bumps in 
their LFs, assuming they are well populated.  Although there is still 
some age/metallicity degeneracy with the RC/HB, RGB bump and AGB bump, 
by analyzing the brightness of each of these features relative to each 
other and relative to the TRGB, we can eliminate any dependence on 
distance and reddening and place considerable constraints on the age and 
metallicity of the stellar population.  Then, with age and metallicity 
in hand, we can use our synthetic CMDs to compare the predicted 
intrinsic color of the RGB with the observed color and thereby determine 
the reddening of the stellar population.  At known reddening, the TRGB 
magnitude provides the galactic distance.

\subsection{Age and Metallicity}
\label{agemet}

\begin{figure}
\vspace{-1.0truecm}
\plotone{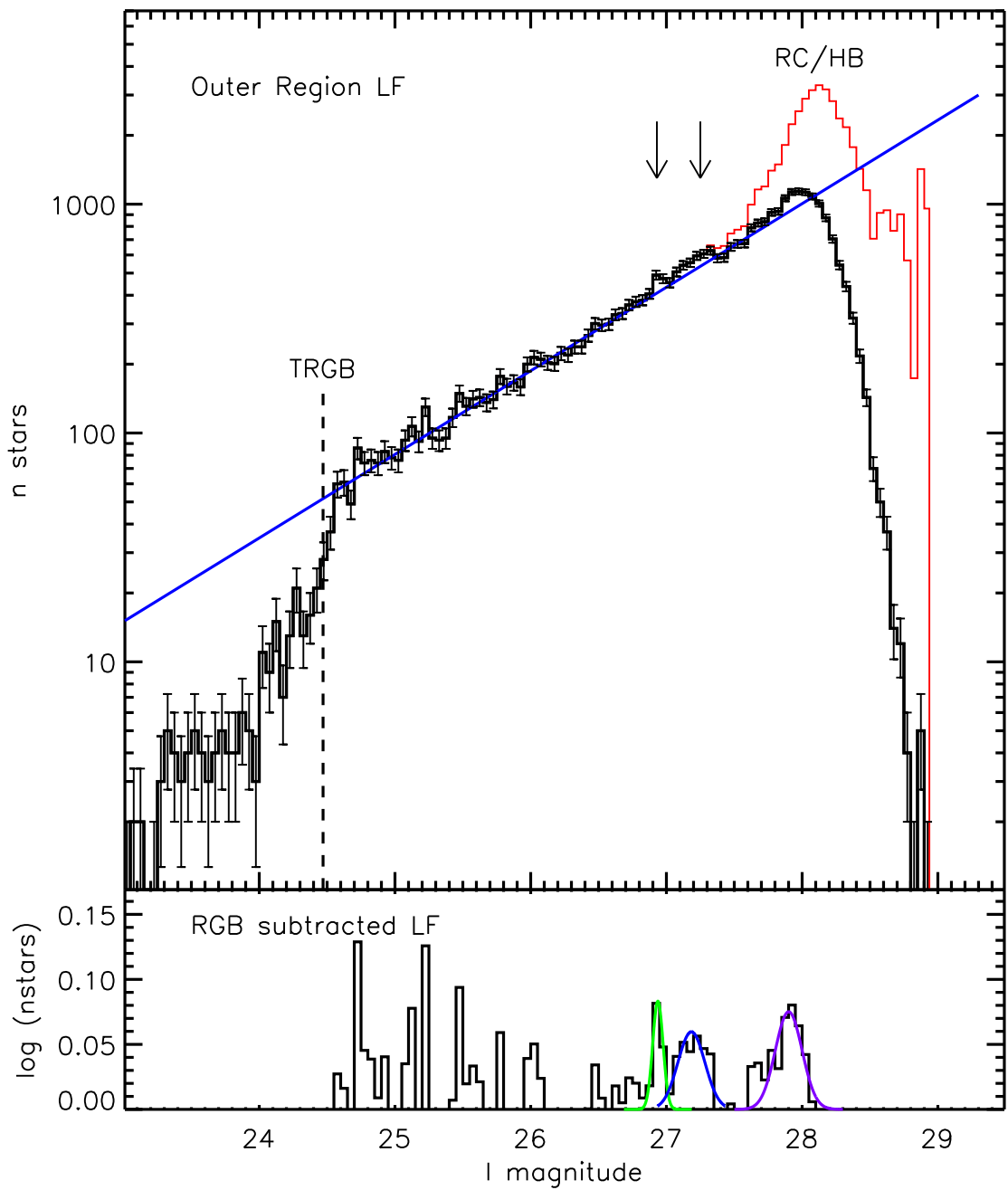}
\vspace{-1.0truecm}
\caption{Observed (black line) and completeness corrected 
(thin red line) LFs of NGC
1569's outer region.  Brighter than I $\sim$ 27.3, our photometry of NGC
1569's outer region is $\sim$ 100\% complete. At $I = 28.4$ the 
completeness
is only 20\%, and the completeness correction is not very reliable
beyond this magnitude. The vertical dashed line marks the position of
the TRGB and the solid (blue) line follows the slope of the RGB. Three
features of interest, the RC/HB (labeled) and the possible AGB bump
and RGB bump features (arrows) are seen to extend above the expected
RGB slope.  Error bars overplotted on the observed LF show the Poisson
noise in each magnitude bin and suggest that the possible AGB bump and
RGB bump features may be real.  The bottom panel shows the RGB 
slope-subtracted LF, with gaussian fits to the three features overplotted.  
}
\label{lf_err}
\end{figure}

In Fig.~\ref{lf_err} we plot the observed LF of NGC 1569's outer region 
as the 
solid black histogram with the completeness corrected LF plotted in red, 
where we have used our artificial star tests to determine the correction 
factor for each magnitude bin.  Fainter than the TRGB (dashed vertical 
line), the LF of NGC 1569's outer region shows a number of peaks, the 
most 
obvious of which is the RC/HB feature at $I \sim$ 28. A comparison 
with the completeness corrected LF shows that we have only reached the 
bright end of the RC/HB, a result of the fact that NGC 1569 is 50\% 
farther away than previously believed (\S 5.1). To determine which of 
the other peaks may be real features, in Fig.~\ref{lf_err} we fit the 
slope of the RGB between $I$ = 24.8 and 26.8 with a solid line.  We also 
show error bars with the Poisson error for each magnitude bin.  
Although none of the other peaks along our LF are as conspicuous as seen 
by other authors (e.g., \citealt{rejkubaetal2005}), there are two small 
bumps, marked with the arrows, that stick up above the RGB and are in 
the correct locations to possibly be the AGB and RGB bumps.

Having identified the RC/HB, as well as AGB and RGB bump candidates in 
NGC 1569's LF, we can now compare the positions of these three features, 
plus that of the TRGB, to the predictions of theoretical isochrones.  To 
measure the magnitudes of both bumps and the RC/HB in the observed LF of 
NGC 1569, we first subtract the slope of the RGB, leaving just the 
features in which we are interested.  We fit gaussians to both bump 
features and take the center of the gaussian as the $I$-band magnitude 
of each feature.  For the two bumps we find: $I_{bump1} = 26.94 \pm 
0.02$ and $I_{bump2} = 27.12 \pm 0.03$, where the errors listed are the 
errors in the fit.  Due to completeness issues near the RC/HB, which 
result in a steep drop off in the LF, fitting a gaussian to the RC/HB 
feature may yield a biased measurement.  Instead, we simply take the 
magnitude of the LF bin with the most stars as our measurement of the 
RC/HB.  For the RC/HB, which is $\sim$ 0.4 mag wide (full width, zero 
intensity), or about eight times the LF bin size, we measure $I_{RC/HB} 
= 27.93 \pm 0.025$.  This is an upper limit on the brightness of the 
RC/HB due to completeness effects.  The quoted error is half the width 
of the magnitude bins in the LF.  In \S\ref{dist} we discuss the 
measurement of the TRGB, for which we find $I_{TRGB} = 24.47 \pm 0.04$.

Due to the fact that our observations only reach down to the RC/HB 
branch feature, and not well below it, completeness and photometric 
errors may play a role in the magnitudes that we measure for the RC/HB 
and RGB and AGB bumps. Therefore, we cannot simply compare our numbers 
to those predicted by stellar evolution models. Instead, we use the same 
code as described below in Appendix~A to create synthetic CMDs from each 
individual isochrone.  In this, we make use of our artificial star test 
results to apply the appropriate errors and incompleteness.  We have 
assumed a reddening of \ebv = 0.56 \citep{israel1988}, which, when 
combined with our measured TRGB value, gives a distance of $\sim$ 3 
Mpc. Then, in the same way as for NGC 1569's LF, we measure the 
magnitudes of the RGB bump, AGB bump, and RC/HB.  At the brightness of 
the TRGB, completeness and errors are not expected to significantly 
affect our measurements.  We therefore use the TRGB magnitude as 
reported in the isochrones.

\begin{figure}
\plotone{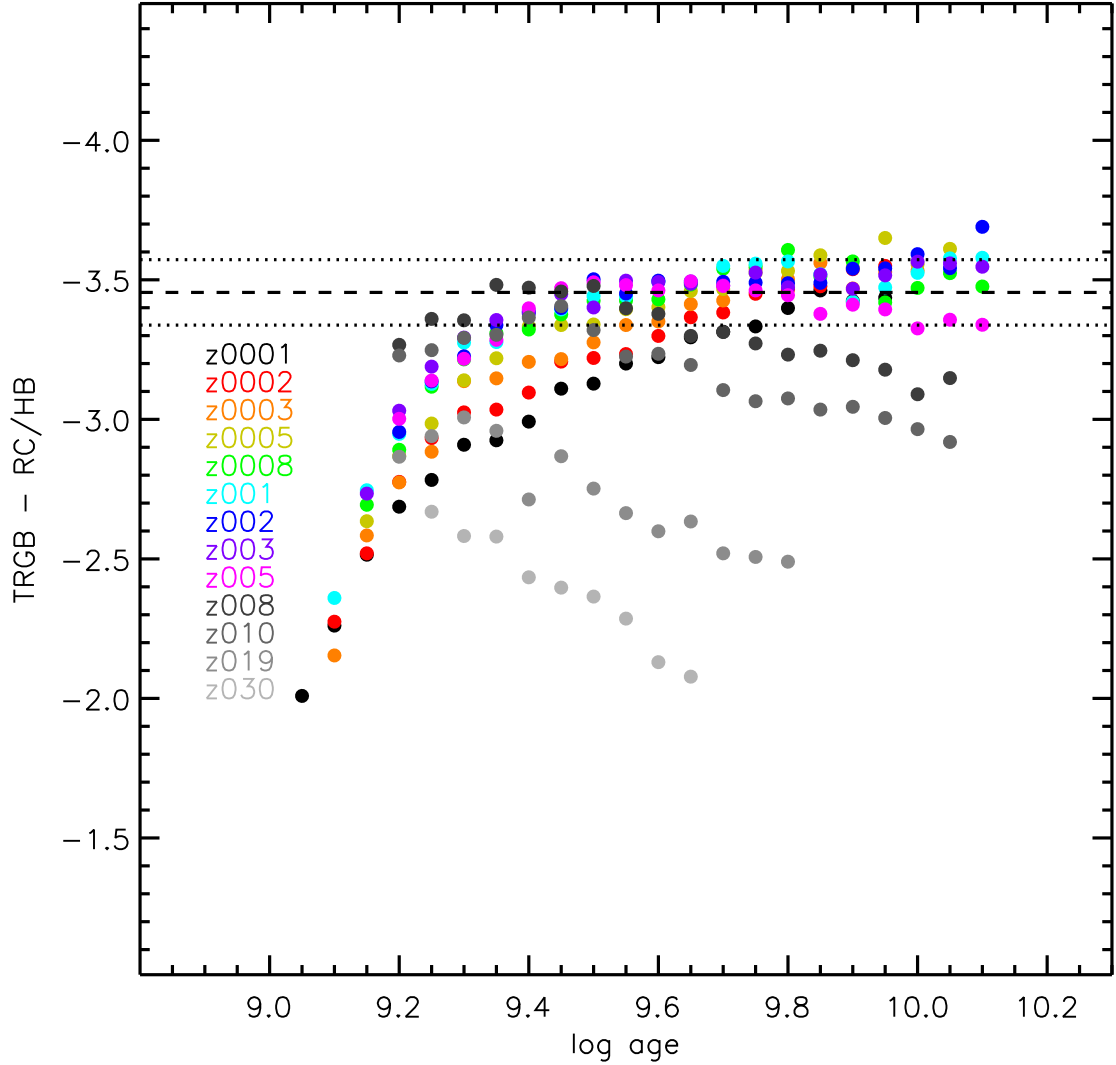}
\caption{Difference in $I$-band magnitude of the TRGB and RC/HB as a 
function of age.  Theoretical values for a range of metallicities 
(shown as symbols of different colors) were calculated by creating 
a synthetic CMD for each isochrone using our SFH code, which takes into 
account errors and incompleteness, and measuring the brightness of the 
features in the same way as for NGC 1569.  The dashed line is our 
measured TRGB-RC/HB value for 
stars in NGC 1569's outer region while the dotted lines mark the 
3$\sigma$ 
error in our measurement.  Only those isochrones with TRGB-RC/HB values 
within 3$\sigma$ of our measured value are considered as possible 
matches to NGC 1569.
}
\label{trgbrchb}
\end{figure}

Since then we are confident in our identification and measurement of the 
RC/HB, we use this feature to make the first cut in determining which 
age and metallicity best fits NGC 1569's outer region.  In 
Fig.~\ref{trgbrchb} 
we plot the values of $I_{TRGB} - I_{RC/HB}$, as predicted by our 
synthetic CMDs, as a function of age for a range of abundances.  
Overplotted as the dashed line is our measured value for NGC 1569, with 
the dotted lines representing the 3$\sigma$ error bars.  Only those 
synthetic CMDs that fall within 3$\sigma$ of our observed value are 
considered further.  Because we have no {\it a priori} knowledge of the 
order of the AGB and RGB bumps, or if the bumps we have identified are 
actually real features, we consider all combinations of these two 
features in determining our best fit LF.  We find that all of the best 
fit models for the RC/HB show only the AGB bump, with the RGB bump 
`lost' in the RC/HB feature. The isochrone that provides the best fit to 
the positions of the TRGB, AGB bump, and RC/HB to within the measurement 
errors has log(age) = 10 (10.0 Gyr) and [Fe/H] = -1.4 (Z = 0.0008).  Its 
LF is plotted in Fig.~\ref{best_lf} as the red histogram along with NGC 
1569's outer region (black histogram).  

While this is the best fit model, there are two major differences 
between the observational data and model.  First, the AGB bump and RC/HB 
features in the model are much more pronounced than what we observe. 
Second, the slope of the synthetic LF is similar to what is seen in star 
clusters ($N(L) \propto L^{-\beta}$ with $\beta \approx 0.32$; e.g., 
\citealt{zoccaliandpiotto2000}) whereas the slope of NGC 1569's RGB is 
much steeper ($\beta = 0.37$).  The poor fit to the bump features 
opens the possibility that the bumps we have identified in the LF in 
Fig.~\ref{lf_err} may not be associated with the actual RGB and AGB 
bumps in NGC 1569.  That is not to say that they do not exist in NGC 
1569, but rather that they have been smoothed out due to a superposition 
of mixed stellar populations.  Combined with the difference between 
predicted and observed LF slopes, this suggests that NGC 1569's LF as a 
whole is not represented by any single age, single metallicity 
population; it must have had a more complex SFH, which we derive 
explicitly in \S\ref{sfh}.

\subsection{Distance and Reddening}
\label{dist}

As discussed in \citet{grocholskietal2008}, our data represent the first 
unequivocal detection of the TRGB in NGC 1569 and thereby allow us to 
determine its distance via the luminosity of the TRGB.  Although we 
presented the distance to NGC 1569 in \citet{grocholskietal2008}, since 
we have reprocessed our data using the most recent HST ACS calibration 
data we revisit our distance calculation, which also depends on the 
reddening.  We calculate the distance and reddening here, still 
assuming NGC 1569 can be represented by a SSP, and discuss the results 
from more complicated SFHs in \S 5.2.

In the LF of a stellar population the TRGB acts as a discontinuity, the 
position of which is easily measured.  Using the software developed by 
one of us (R.~P.~v.~d.~M.) and detailed in \citet{cionietal2000}, we 
find that $I_{TRGB} = 24.47 \pm 0.04.$ The location of the TRGB is 
marked by the vertical dashed line in Fig.~\ref{lf_err}.  We note that 
this value is slightly fainter than reported in 
\citet{grocholskietal2008} due to the updated processing and 
photometry.

For the reddening appropriate for NGC 1569, we can either use values 
from the literature (as we did in \citealt{grocholskietal2008}), or we 
can estimate it from our data.  Literature values for the foreground 
extinction toward, and intrinsic reddening in, NGC 1569 span a wide 
range of values.  The most commonly adopted value is \ebv = 0.56 
\citep{israel1988}, which was calculated using integrated UV photometry 
of the core of NGC 1569.  Since the recent star formation, and therefore 
the gas and dust, in NGC 1569 is concentrated in the core, we expect the 
reddening in the outer region to be lower.  Thus, in 
\citet{grocholskietal2008} 
we adopted the foreground extinction due to the Milky Way provided by 
the \citet{bh1982} reddening map, \ebv = 0.50.  We note that the 
reddening map of \citet{sfd1998} gives a much higher value, \ebv = 0.68, 
for NGC 1569.  However, as discussed by \citet{relanoetal2006}, this 
estimate of the {\it foreground} extinction is higher than the {\it 
total} extinction in NGC 1569 derived by many authors 
(e.g.~\citealt{israel1988}, \citealt{origliaetal2001}) suggesting that 
it is likely an overestimate of the reddening at the position of NGC 
1569.

In the previous section we used the relative brightnesses of the TRGB, 
AGB bump, and RC/HB features to determine an estimate for the age and 
[Fe/H] of the dominant population in NGC 1569.  Since the RGB spans a 
very small range in color, this approach gives an age and metallicity 
that is independent of the reddening of the stellar population.  Thus, 
to determine \ebv from our data we can compare the observed RGB color of 
NGC 1569 with that predicted for a synthetic population that has an age 
of 10 Gyr and [Fe/H] = -1.4.  For RGB stars on the synthetic CMD $\sim$ 
0.5 mag below the TRGB, we find an intrinsic color of $(V-I)_0 = 1.29$.  
Over the same magnitude range, RGB stars in NGC 1569's outer region have 
an 
average apparent color $(V-I) = 1.98$.  This implies a reddening of \ebv 
= 0.58 (adopting the reddening law of \citet{ccm1989}, where $A_V = 3.1 
\ebv$ and $A_I = 0.614 A_V$ for photometry on the Johnson-Cousins 
system).

The absolute $I$-band magnitude of the TRGB ($M_I^{TRGB}$) is well known 
to be roughly constant around $\sim$ -4.0 mag in SSPs with age 
$\gea$ 2 Gyr and [Fe/H] $\lea$ -0.5. \citet{barkeretal2004} used 
synthetic CMDs to study the reliability of the TRGB as a standard candle 
for resolved stellar populations with complex SFHs.  They found that 
$M_I^{TRGB} = -4.0 \pm 0.1$ independent of the SFH, provided that the 
median dereddened color of RGB stars $\sim$ 0.5 mag below the TRGB is 
$(V-I)_0 \la 1.9$.  It is therefore appropriate to assume $M_I^{TRGB} = 
-4.0 \pm 0.1$ for our distance calculations.  Combined with $I_{TRGB} = 
24.47 \pm 0.04$ and $\ebv = 0.50 \pm 0.05$ (from \citealt{bh1982}), this 
yields $\mmo = 27.52 \pm 0.14$ or $D = 3.19 \pm 0.21$ Mpc.  If we 
instead use the reddening estimated from our RGB analysis, $\ebv = 0.58 
\pm 0.06$ (we assume a 10\% error), we find $\mmo = 27.36 \pm 0.16$ or 
$D = 2.96 \pm 0.22$ Mpc.  Both of these distances are much farther than 
the typically assumed distance of 2.2 Mpc (e.g., \citealt{israel1988}) 
and place NGC 1569 on the near edge of the IC 342 group of galaxies.

\section{Star Formation History Extraction from the CMD}
\label{sfh}

\subsection{Methodology}
\label{sfhmethod}

We showed in \S\ref{LFs} that the LF of the NGC 1569 outer region is not 
well
fit by an SSP and must have had a more complex SFH. To infer this SFH
we use the new SFHMATRIX code (van der Marel \& Grocholski in prep.)
described in Appendix A. Starting from a set of isochrones, and for
any assumed distance and extinction, the code finds the SFH as
function of age and metallicity that best matches the Hess diagram of
the observed CMD in a $\chi^2$ sense.  We note that the full SFH 
calculations are independent of our SSP analysis and that the accuracy 
of our SFH is ultimately tied to the accuracy of the stellar evolution 
models we have adopted.

For a given assumed extinction, we find the best-fitting distance as 
follows. We start with a trial distance and find the best fitting SFH. 
From this we create a synthetic CMD realization (i.e., a Monte-Carlo 
realization that has the same number of stars on the CMD, taking 
incompleteness and photometric errors into account, as the observed 
CMD). We then analyze both the observed and the synthetic CMD with the 
TRGB analysis software described in Section~\ref{dist}. Based on the 
inferred difference between the observed and synthetic TRGB magnitudes 
we adjust the trial distance, and iterate this procedure till 
convergence (agreement better than $0.01$ mag). The iteration is 
necessary because not all SFHs yield the same absolute TRGB magnitude; 
the distance depends on the SFH. There generally is agreement at the 
$\pm 0.1$ mag level \citep[e.g.,][]{barkeretal2004}, but this is not 
sufficient for the most accurate results.

To infer the extinction, we run the procedure for a range of trial 
extinction values and plot the $\chi^2$ of the CMD fit as a function of 
extinction (see Fig.~\ref{chisq}). The data vary smoothly and are well 
fit by a fourth-order polynomial (dashed curve). This has a minimum at 
\ebv = 0.58, which provides the best-fitting extinction. The 
corresponding best-fitting distance modulus is \mmo = 27.43, yielding a 
distance of 3.06 Mpc. These results agree very well with published 
reddenings as well as the results from our SSP analysis of the LF and 
RGB color in Section~\ref{dist}. In principle, the random errors on the 
extinction and distance can be robustly determined using a repeated 
analysis of pseudo-data created from the observed CMD data set using 
bootstrapping (see Appendix A). In practice, systematic errors of 
various kinds are probably of more importance. Based on the combined 
insights from literature studies of the extinction, and our own SSP and 
SFH analyses, we adopt as our final estimates $\ebv = 0.58 \pm 0.03$ and 
$D = 3.06 \pm 0.18$ Mpc. The distance error is dominated by the 
systematic uncertainty in the knowledge/calibration of the absolute 
magnitude of the TRGB feature.

It is of interest to note that the SFH analysis yields extinction values 
that are consistent with those inferred through independent techniques. 
This means that the CMD data has enough information content to break the 
age-metallicity-extinction degeneracy for old populations.  The 
color and curvature of the RGB is driven primarily by metallicity with 
only a small contribution from age.  Metal-poor isochrones are 
relatively blue with little curvature, while metal-rich isochrones are 
red and have a significant amount of curvature.  Therefore, if we use 
too low of a value for the extinction, the SFH code must use the 
metal-rich isochrones to match the observed color of the RGB, resulting 
in a synthetic CMD with too much curvature on the RGB.  Conversely, high 
values of extinction lead to metal-poor synthetic CMDs with too little 
RGB curvature.  Since our data reach $\sim 4$ mag below the TRGB, the 
curvature is observationally well constrained. We found that the high 
$\chi^2$ values in Fig.~\ref{chisq} for models with relatively low or 
high extinction values are driven in part by the fact the models for 
these extinctions cannot fit the observed RGB curvature at the given 
(observed) color.

\subsection{The CMD Fit}

Fig.~\ref{sfh_cmds} compares the observed CMD (right) to a synthetic CMD 
realization (left) from the best-fit SFH. The area below the solid line 
was excluded from the $\chi^2$ minimization because the completeness is 
below 20\% there, making the artificial star corrections unreliable. The 
area above and to the right of the dashed box was also excluded, for two 
reasons. First, this is where foreground stars are found (see 
Fig.~\ref{besancon}), which are not explicitly accounted for in our 
models. Second, this is where thermally pulsating AGB (TP-AGB) stars are 
located. Near the end of their AGB lifetimes, these stars undergo a 
series of He shell flashes due to the He- and H-burning shells turning 
on and off.  While the updated Padova models (see 
\citealt{girardietal2010}) provide a much improved treatment of the 
TP-AGB stars, this is still a complex and highly uncertain phase of 
stellar evolution.  Moreover, a significant but unknown fraction of 
TP-AGB stars may be obscured by dust shells \citep{boyeretal2009} making 
it difficult to use such stars to constrain the SFH.

In the region of the CMD that was fitted, the agreement between the
synthetic and observed CMDs is excellent. The $\chi^2$ of the fit is
2668, for $N_{\rm DF} = N_{\rm pix} - N_{\rm basis} = 898$ degrees of
freedom. Here $N_{\rm pix} = 1773$ is the number of pixels in the CMD
Hess diagram that was fitted, and $N_{\rm basis} = 875$ is the number
of different (isochrone) basis functions that was used to build the
fit. The ratio $\chi^2/N_{\rm DF} = 2.97$ is somewhat higher than
expected purely from random errors, but such ratios are not atypical
for studies of this kind \citep{mcquinnetal2010}. The only visually
obvious discrepancies between the synthetic and observed CMDs occur in
the regions that were masked in the fit. These discrepancies are
well-understood: shortcomings in the TP-AGB evolutionary models at
bright magnitudes and shortcomings in the correcting of very
incomplete data with artificial start tests at faint magnitudes. These
do not impact the inferred SFH, because of the exclusion of these
regions from the fit.

\begin{figure}
\plotone{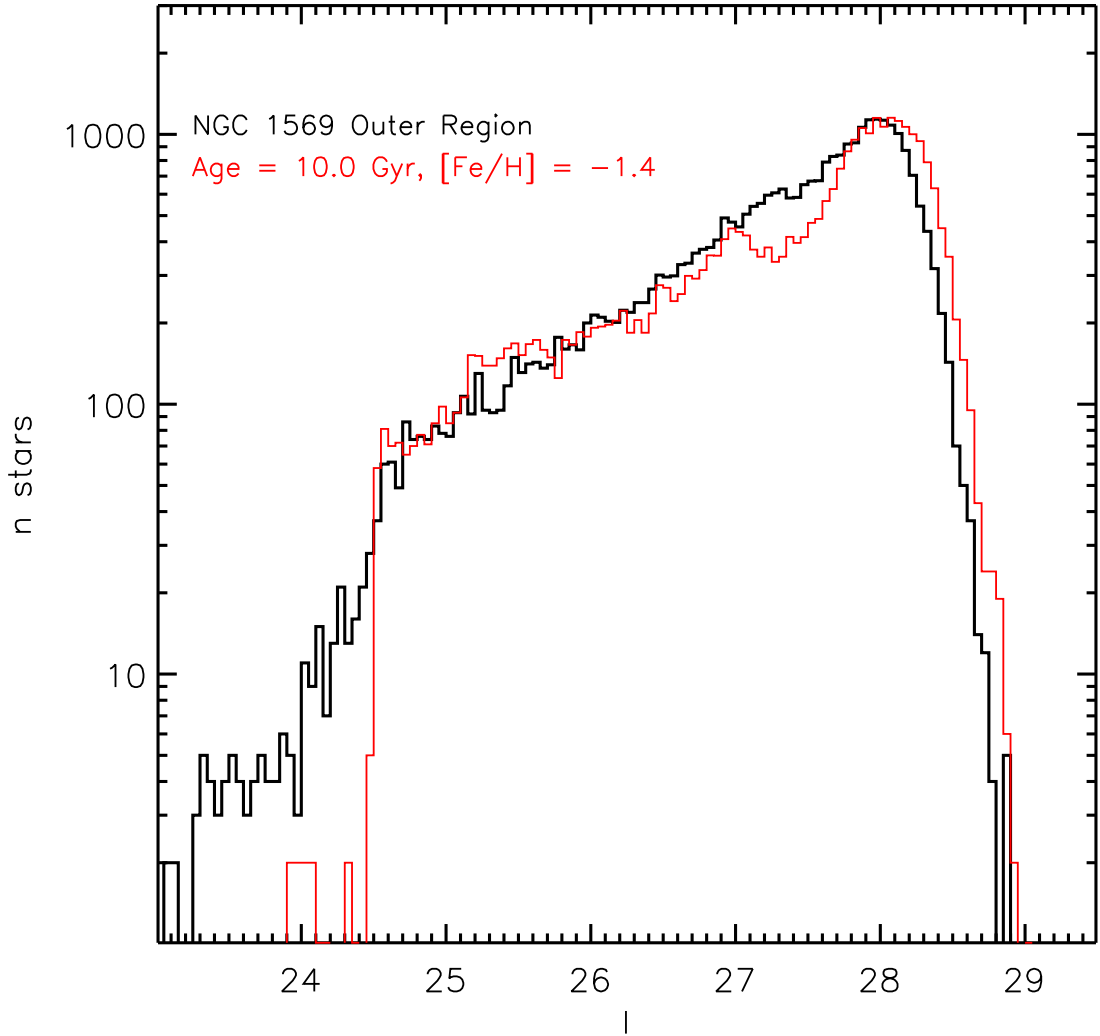}
\caption{Theoretical SSP LF (thin red line) compared to NGC 1569's outer 
region 
LF (black line).  The synthetic LF shown, with an age of 10 Gyr and 
[Fe/H] = -1.4, provides the best simultaneous match to the positions of 
the AGB bump and RC/HB relative to the TRGB, to within our measurement 
errors.  At this age and metallicity, the RGB bump is faint enough that 
it blends in with the RC/HB.  While this synthetic population provides 
the best match, its LF features are much more pronounced and its LF 
slope flatter than what we observe.  This difference suggests that NGC 
1569's outer region is not an SSP.  Instead, it must have undergone many 
epochs 
of star formation, which served to both smooth the LF features and make 
it steeper.
}
\label{best_lf}
\end{figure}

\begin{figure}
\plotone{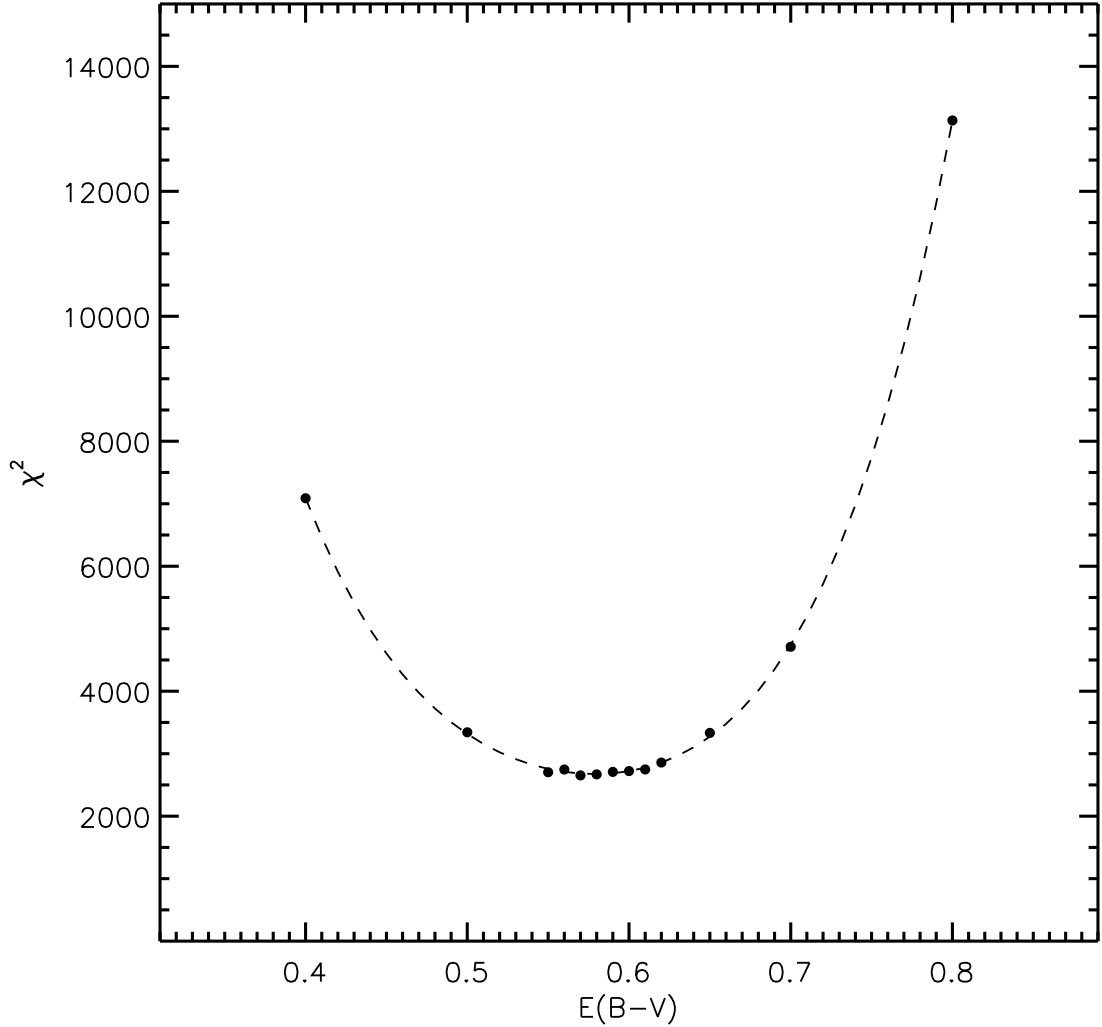}
\caption{$\chi^2$ of our SFH model fits to the CMD as a function of
reddening.  The $\chi^2$ values vary smoothly with \ebv and are well
fit with a 4th order polynomial (dashed curve).  The fit has a minimum
at \ebv = 0.58. At this best-fit reddening, the galaxy distance
implied by the TRGB magnitude is $3.06$ Mpc.}
\label{chisq}
\end{figure}

\begin{figure}
\plotone{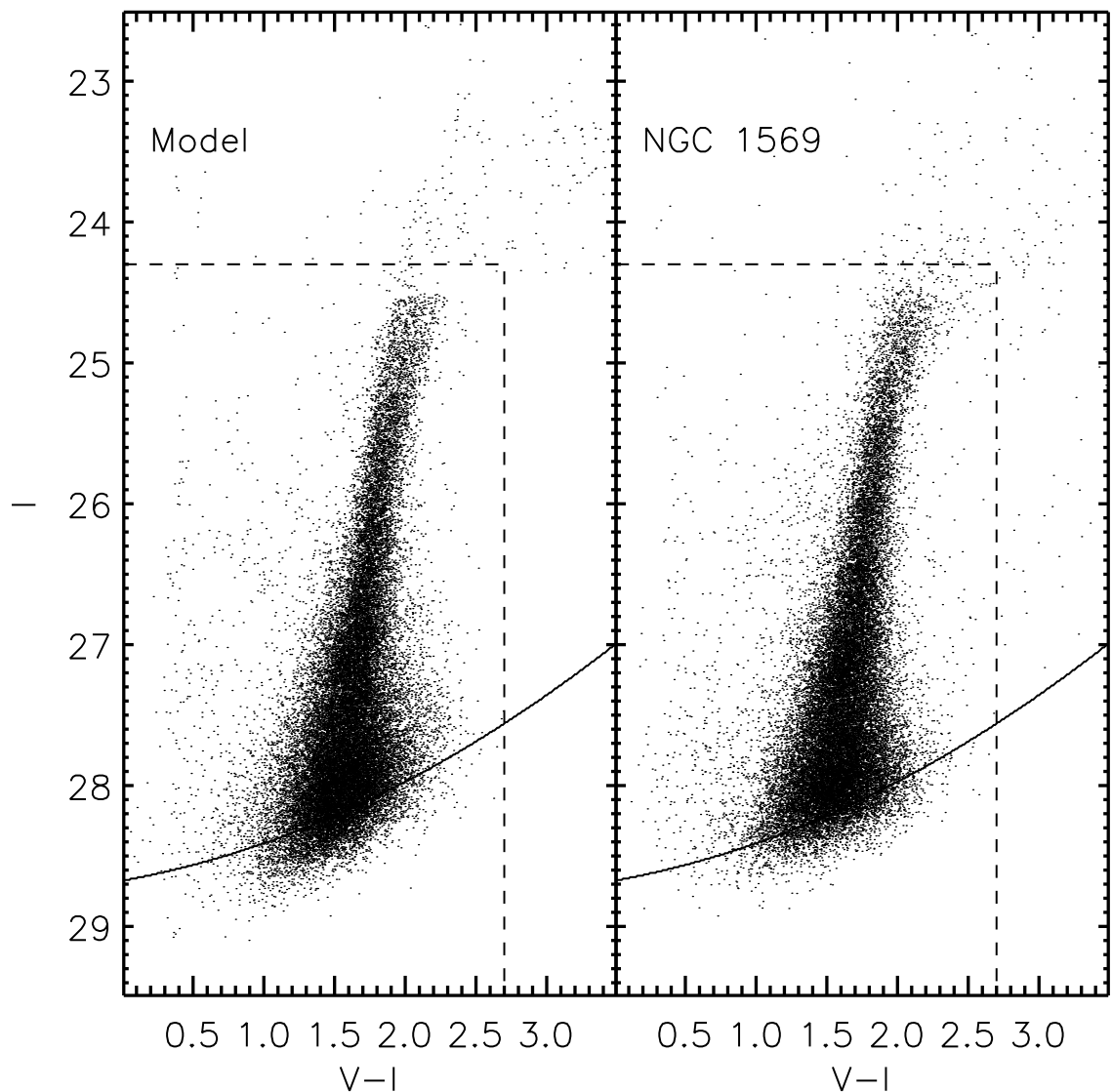}
\caption{Comparison of the observed CMD (right) to a synthetic CMD
realization (left) from the best-fit SFH. The areas below the solid
line and above and to the right of the dashed box were excluded from
the fit. At the faint magnitudes the completeness is too low ($<20$\%)
to yield an accurate fit.  At the bright magnitudes there are
foreground stars (see Fig.~\ref{besancon}) and TP-AGB stars, a complex 
state of stellar evolution that has historically been difficult to 
model.  The model matches the data well in the RGB region that
was fit.}
\label{sfh_cmds}
\end{figure}

\begin{figure}
\plotone{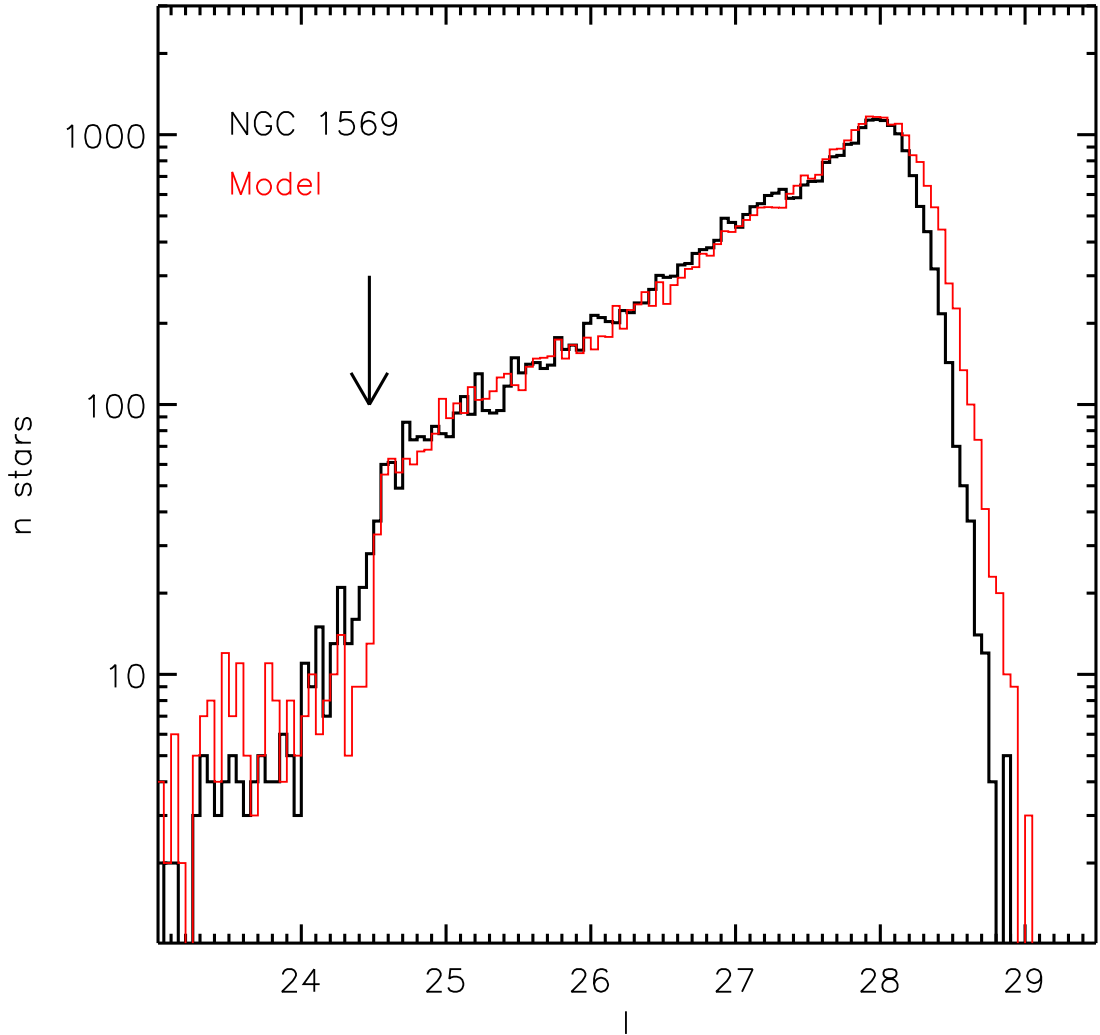}
\caption{Luminosity function of NGC 1569 (black line) compared to the 
model LF for the best-fit SFH (thin red line). The TRGB is marked with 
the arrow.  The 
isochrones slightly overpredict the number of TP-AGB stars brighter than 
$I \sim 24$.  Below the TRGB, the model provides excellent agreement 
with the data. Discrepancies at $I \gtrsim 28$ are due to the 
limitations of artificial star test corrections when the incompleteness 
gets very low.  
}
\label{sfh_lf}
\end{figure}

Fig.~\ref{sfh_lf} compares the LF of NGC 1569 ({\it black}) to the
model LF for the best-fit SFH ({\it red}). As described above, there
are well-understood discrepancies brighter than the TRGB (arrow) and
fainter than $I \gtrsim 28$. However, at the intermediate magnitudes
that were actually included in the $\chi^2$ minimization, the fit is
excellent. This is particularly true when compared to the predictions
of the best-fit SSP in Fig.~\ref{best_lf}. The best-fit SFH shows
excellent agreement with the observed slope of the RGB, the observed
magnitude and prominence of the RC/HB, and the lack of other prominent
features (e.g., the RGB and AGB bumps).  

Fig.~\ref{sfh_lf} compares the LF of NGC 1569 ({\it black}) to the model 
LF for the best-fit SFH ({\it red}). As described above, there are 
well-understood discrepancies brighter than the TRGB (arrow) and fainter 
than $I \gtrsim 28$. However, at the intermediate magnitudes that were 
actually included in the $\chi^2$ minimization, the fit is excellent.  
The best-fit SFH matches the observed slope of the RGB, the 
observed magnitude and prominence of the RC/HB, and the lack of other 
prominent features (e.g., the RGB and AGB bumps).  This is in sharp 
contrast to our best fit SSP in Fig.~\ref{best_lf}, and confirms our 
assessment in \S ~\ref{agemet} that NGC 1569 is not well fit by any SSP 
and must have had a more complex SFH.

\subsection{Star Formation History}
\begin{figure}
\plotone{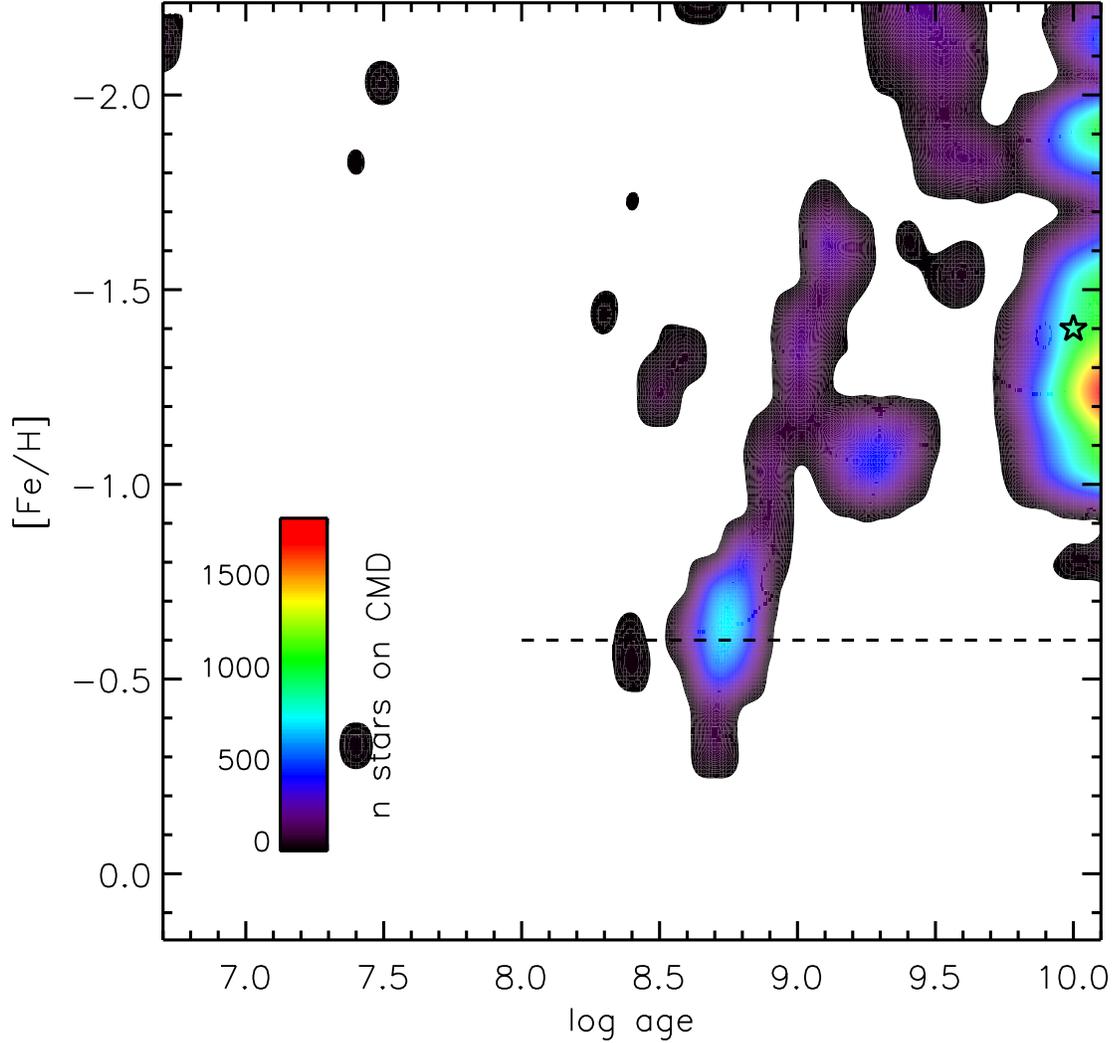}
\caption{Contour plot of the SFH of the NGC 1569 outer region, 
interpolated
onto a grid with a step size of 0.01 in both log (age) and [Fe/H]. The
color coding indicates the predicted number of stars on the parts of
the CMD that are not masked in Fig.~\ref{sfh_cmds}. The open star marks
the position of the best-fitting SSP model as derived in
\S\ref{agemet}. The dashed line indicates the abundance of the gas in
NGC 1569. The properties and interpretation of this SFH are discussed
in \S\ref{sfh}.  
}
\label{sfh_contour}
\end{figure}

Fig.~\ref{sfh_contour} shows the full SFH of NGC 1569's outer region as 
a color-contour plot as a function of log (age) and [Fe/H]. The quantity 
that is plotted is the predicted number of stars on the unmasked portion 
of the CMD. Thus, this provides a direct assessment of which 
(age,metallicity) combinations contribute most to the observed CMD. The 
SFH shows a number of salient features, which we will discuss in turn.

{\bf (1)} The peak of the SFH is found at old ages (log(age) $\gtrsim 
10$) and metallicities from [Fe/H]$ = -1$ to $-2$, but centered around 
[Fe/H]$ = -1.25 \pm 0.1$.  This is very similar to the properties of the 
best-fit SSP inferred in \S\ref{agemet}, which has log(age) $= 10.0$ and 
[Fe/H]$ = -1.4$ (open star in Fig.~\ref{sfh_contour}).  There is a 
well-known age-metallicity degeneracy in the modeling of old stellar 
populations. Before accepting a SFH result, one must therefore ask if 
the age-metallicity dependence of the inferred SFH is uniquely implied 
by the data, and if the data did indeed have the information content 
necessary to ameliorate the degeneracy. In this case, the answer is yes.  
We have tried to match our observations of NGC 1569 using basis 
functions with restricted parameter space [e.g., log (age) $\leq$ 9.9 
(8 Gyr)], but find that with these restrictions we cannot match the 
observed LF.  Thus, while the precise age and metallicity of stellar 
populations can only be determined through a combination of spectroscopy 
and MSTO photometry, the information available on the RGB (brightness of 
the TRGB and RC/HB, color and curvature of the RGB) is sufficient to 
prefer a predominantly old population with a spread in metallicity.

{\bf (2)} The SFH shows a general trend of younger stars having higher 
metallicities. The metallicity of the gas in the central regions of NGC 
1569 has been measured to be [Fe/H]$ = -0.6$ (dashed horizontal line in 
Fig.~\ref{sfh_contour}). There are few stars in the SFH model at 
metallicities higher than this, and the youngest stars inferred with 
significance (log(age) $\approx 8.7$) have a metallicity that approaches 
this value. This is all as generically expected in scenarios of chemical 
evolution. Younger stars formed from gas that was pre-enriched by the 
ejecta of older stars, and the youngest stars should have metallicities 
similar to that of the gas. These results therefore also provide 
additional credibility for the SFH results.

\begin{figure}
\plotone{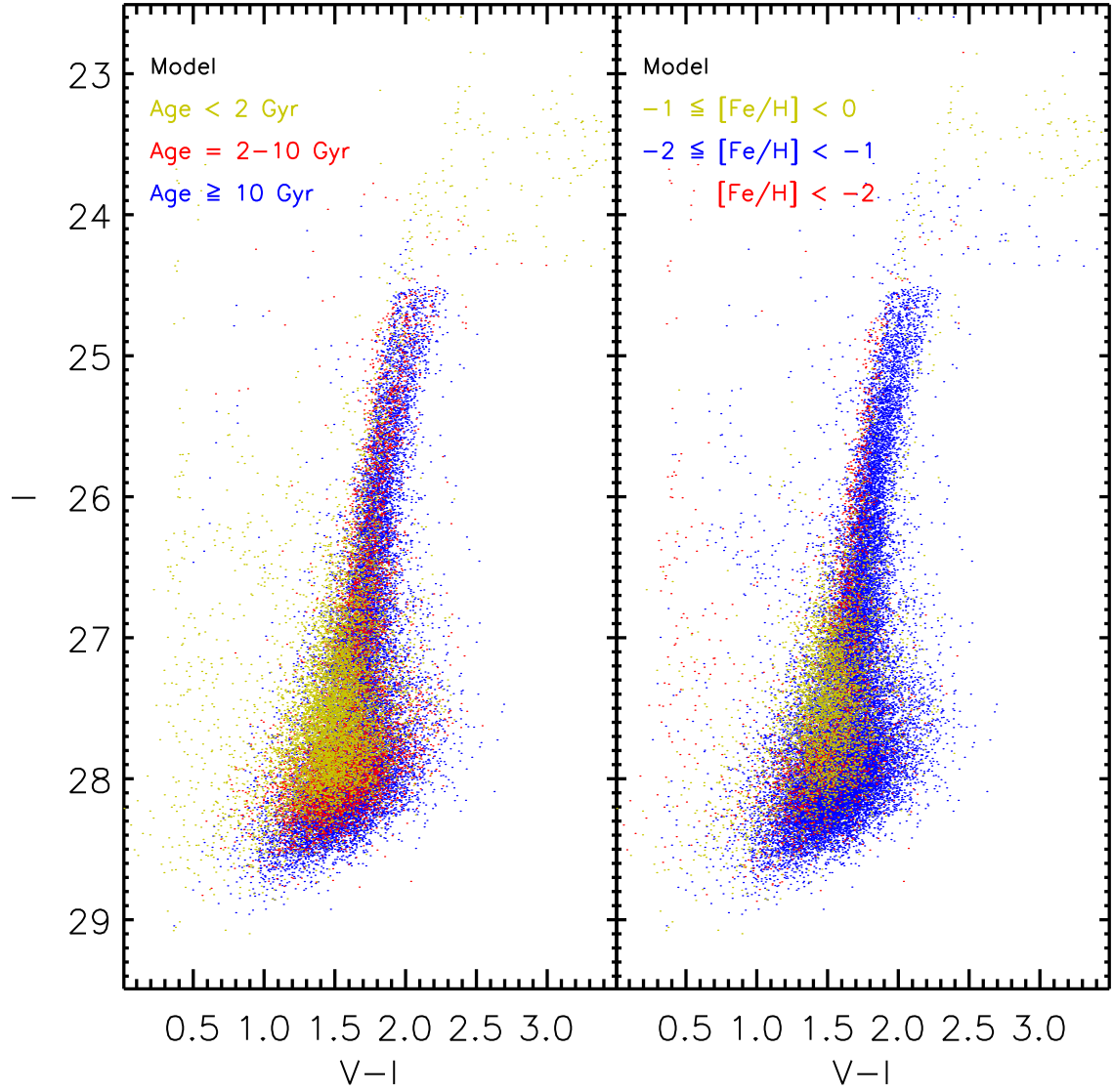}
\caption{A synthetic CMD realization from the best-fit SFH.  Different
colors indicate a break-down into separate age ({\it left}) and
metallicity ({\it right}) components. Stars in yellow correspond to a
young and metal-rich component that steepens the RGB LF at the faint
end, in accordance with the observations.}
\label{sfh_breakdown}
\end{figure}

{\bf (3)} Even though the CMD only shows a predominant RGB, the SFH 
actually has significant star formation in the outer region for the age 
range 
from approximately $0.5-2.0$ Gyr ago (log(age) between $8.7$ and $9.3$). 
Since this may appear counter-intuitive, it is important to understand 
why this occurs in the models.  The luminosity at the TRGB, i.e. the 
point of central He ignition, is virtually constant with increasing 
stellar mass, up to close to the {\it RGB phase transition} mass.  Stars 
that are less massive than the phase transition mass have fully 
degenerate cores when they begin He fusion, while He ignition occurs in 
more massive stars under non-degenerate conditions 
\citep{sweigartetal1990}.  As stars approach this transition mass, the 
luminosity of the TRGB rapidly decreases by $\sim$ 2 mag in the 
$I$-band.  In terms of age, the RGB phase transition occurs in the 
Padova isochrones at $\sim$ 1-2 Gyr, with the younger, more massive 
stars populating a short RGB, while the older, less massive stars form 
the fully extended RGB.  Our observed RGB luminosity function can only 
be reproduced with an intermediate age component, which steepens the LF 
by populating the magnitude range $27 \la I \la 28$, on top of the 
extended RGB component provided by the older stellar populations.  This 
argument is illustrated in Fig.~\ref{sfh_breakdown} where we plot the 
synthetic CMD realization from the best-fit SFH, broken down into its 
age ({\it left}) and metallicity ({\it right}) components.


{\bf (4)} The {\it observed stars} in the outer region span a range of 
ages,
from $\sim 0.5$ Gyr to a Hubble time (Fig.~\ref{sfh_contour}). However,
the {\it mass} in the outer region resides almost exclusively in the 
oldest
stars. As a population ages, the fraction of the stars that is
observable decreases. Therefore, a single observed star that is old
hints at a much larger underlying reservoir of unseen stars than does
a star that is younger. For the best-fit SFH, 93\% of the total outer 
region
mass resides in stars with $\log(age) \geq 9.7$ (age $>5$ Gyr).

\begin{figure}
\plotone{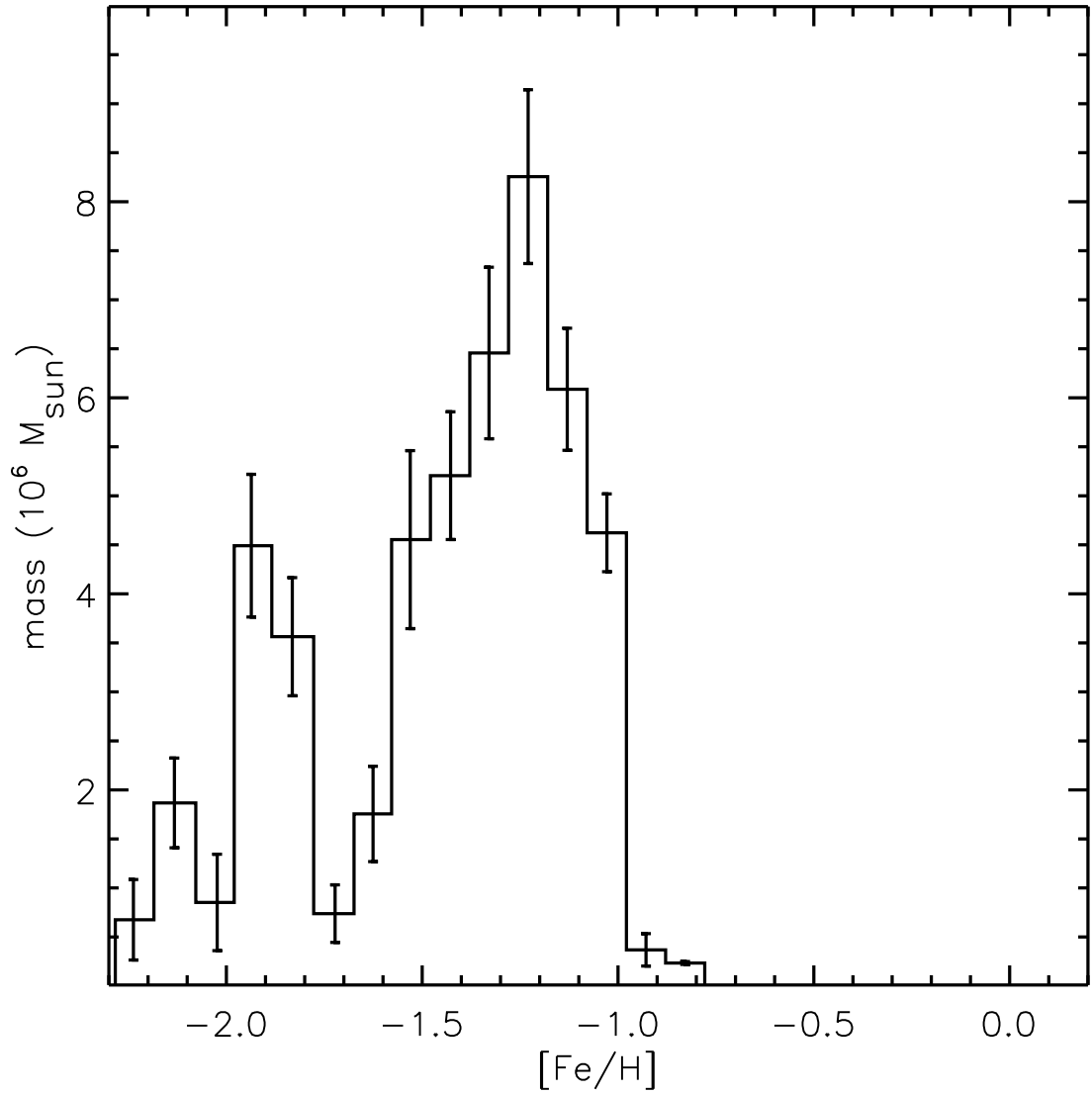}
\caption{Stellar mass in the best-fit SFH vs.~[Fe/H], for those stars
with $\log(age) \geq 9.7$. Error bars were calculated using
bootstrapping as described in Appendix~A. The old stars shown in this 
figure show a
significant spread in metallicity, ranging between [Fe/H]$ = -1$ and
$-2$.}
\label{massvfeh}
\end{figure}

{\bf (5)} Even the oldest stars have a significant spread in 
metallicity, ranging between [Fe/H]$ = -1$ and $-2$. This is evident 
from Fig.~\ref{massvfeh}, which shows the distribution of stellar mass 
vs. [Fe/H], for only those stars with $\log(age) \geq 9.7$. The error 
bars indicate that the individual peaks and valleys seen as a function 
of metallicity in this range are probably not statistically significant, 
with the possible exception of a dip around -1.8 dex. However, the 
spread in metallicity is a persistent feature of the various SFH fits we 
have explored. This is due to the observed RGB width, which exceeds the 
photometric errors, and therefore requires a metallicity spread to be 
reproduced by the SFH code. Note that the total mass in the field under 
study is small, because we are looking at a low density field in the 
outer region. The integral under the histogram in Fig.~\ref{massvfeh} 
corresponds to only $5.0 \times 10^7 M_{\odot}$.  The spread in [Fe/H] 
in NGC 1569's outer region is similar to that of the Milky Way halo; 
using SDSS 
data, \citet{ivezicetal2008} find that the Milky Way halo metallicity 
distribution is well fit by a gaussian with [Fe/H] = -1.46 and an 
intrinsic width, $\sigma$ = 0.30.

\begin{figure}
\plotone{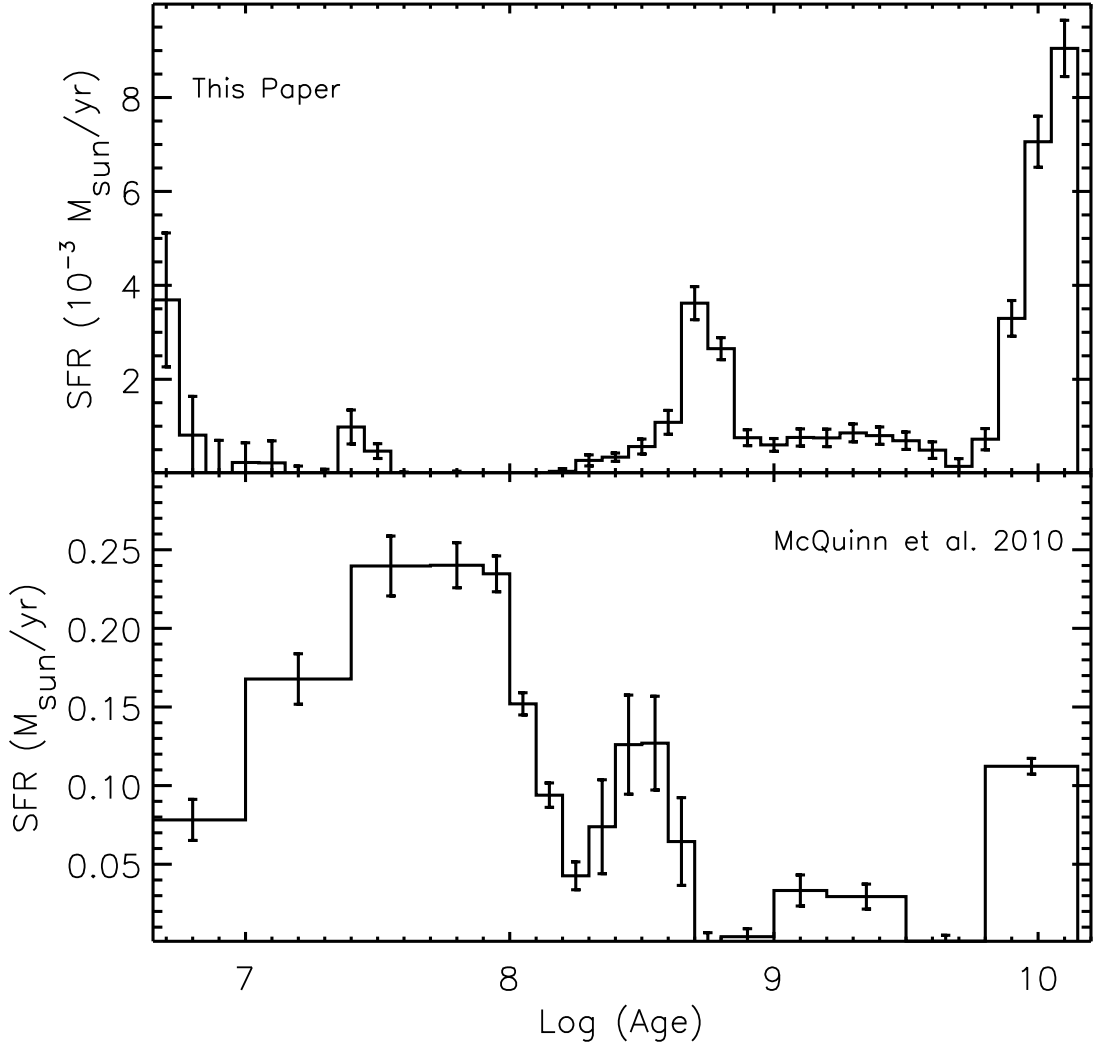}
\caption{SFR as function of log(age), integrated over all
metallicities. ({\it top}) NGC 1569 outer region, as derived in the 
present
paper. ({\it bottom}) NGC 1569 total, heavily dominated by the core,
from \citet{mcquinnetal2010}. The core shows much elevated star
formation at young ages. But for the more ancient populations, our
studies are in good overall agreement: most of the old stars formed
$\sim$ 10 Gyr ago; a low but significant amount of star formation
persisted between 1 and 10 Gyr ago; and there was a SF peak/burst at
$\sim 0.3$--$0.7$ Gyr ago.}
\label{sfrvage}
\end{figure}

{\bf (6)} After most of the outer region stars formed early in the 
Universe ($\log(age) \geq 9.7$), NGC 1569 experienced a relatively 
constant rate of star formation until $\sim$ 0.5 Gyr ago. 
Fig.~\ref{sfrvage} shows the star formation rate (SFR; expressed in 
$M_{\odot}/yr$) as function of log(age) (integrated over all 
metallicities). The error bars indicate that the increase in SFR some 
$\sim 0.5$ Gyr ago ($\log(age) \geq 8.7$) may be statistically 
significant, but not necessarily so.  The oldest epoch of star formation 
in NGC 1569's outer region is congruent to that of the Milky Way 
globular clusters, which formed from the earliest times up until about 7 
Gyr ago \citep{marinfranchetal2009}.  This is also similar to the 
$\Lambda$CDM simulations of a Milky Way-like halo by 
\citet{fontetal2006}, which showed that 80\% of the mass in the inner 
halo (R $<$ 20 kpc) was accreted by $\sim$ 9 Gyr ago, with almost no 
mass ($\ll 1\%$) assembled in the last 5 Gyr.  For reference, at the 
distance of NGC 1569, the full ACS field-of-view is $\sim$ 3.1 kpc 
$\times$ 3.1 kpc. In contrast to the oldest stars, the extended 
formation of stars in NGC 1569's outer region until 0.5 Gyr ago is 
inconsistent with the formation scenarios of the Milky Way halo, 
indicating that NGC 1569 likely had atypical environmental influences 
(see \S\ref{triggers}).

{\bf (7)} The outer region of NGC 1569 does have a few young stars. The
sparsely populated CMD plume at $V-I \approx 0.4$ (see
Fig.~\ref{halo_cmd}) is fit by the SFH code as a low level of star
formation at ages $\lesssim 30$ Myr. It is possible that these stars
may have formed closer to the core of the galaxy, followed by outward
migration through dynamical processes. To travel from the galaxy
center to the outer region in 30 Myr requires a transverse 
velocity of
$\sim 50$ km/s, which is not out of the question.

\subsection{Core-Outer Region SFH Comparison}

We have focused here on a study of the SFH of the NGC 1569 {\it
outer region}. It is of interest to see how our results compare to those
previously derived by other authors for the {\it core} of the galaxy.
The most recent analysis of this was performed by
\citet{mcquinnetal2010}. They used a subset of our ACS images,
performed their own photometry and determined the SFH of NGC 1569 via
CMD fitting with MATCH. Similar to us, they assumed a Salpeter IMF and
used the Padova stellar evolution models. Because of uncertainties due
to the extreme crowding and differential reddening in the central
region of the galaxy, \citet{mcquinnetal2010} required that [Fe/H]
increases with time. While our analysis allows more general chemical
evolution, the inferred SFH for the outer region is in fact generally
consistent with an increasing [Fe/H] with time. The published
\citet{mcquinnetal2010} SFH refers to the entire ACS
field.\footnote{\citet{mcquinnetal2010} divided the galaxy into low
and high surface brightness regions, which is likely similar to our
core/outer region distinction. Their published SFH is the sum of the
individual SFRs for the low- and high-surface brightness regions.}
Since the stars and mass of NGC 1569 are heavily concentrated towards
the galaxy core, that region carries almost all the weight in their
SFH. For simplicity, we'll refer to their result as the SFH of the
core.

In Fig.~\ref{sfrvage} we compare our outer region SFR as a function of 
age (top panel) to the core SFR from \citet[][bottom 
panel]{mcquinnetal2010}. Of course, the core SFR is significantly 
enhanced at young ages compared to what is seen in the outer region. But 
for the more ancient populations, our studies are in good overall 
agreement: most of the old stars formed $\sim$ 10 Gyr ago; a low but 
significant amount of star formation persisted between 1 and 10 Gyr ago; 
and there was a SF peak/burst at $\sim 0.3$--$0.7$ Gyr ago. The 
similarity between the outer region and core SFHs for old stars implies 
that there are no strong radial population gradients for the old stars. 
This is also consistent with our recent study in \citet{rysetal2011} of 
HST/WFPC2 parallel fields much further out in the outer region. Out to 8 
scale radii, we detected no variation in either RGB color or 
carbon-to-RGB star count ratio in the outer region. In view of these 
results, the results for our outer region can be taken to be 
representative of the old stellar populations throughout the galaxy.

As with SFHMATRIX, MATCH can be used to determine the best fit distance 
and reddening of the galaxy.  For the high surface brightness region, 
\citet{mcquinnetal2010} find a best fit distance of 3.2 $\pm$ 0.1 Mpc 
and \ebv = 0.58 $\pm$ 0.03, similar to what we find for the outer region 
(see \S 5.1).  However, for the low surface brightness region, which 
should be similar to our outer region, they find a larger distance, 3.5 
$\pm$ 0.1 Mpc, and lower reddening, \ebv = 0.48 $\pm$ 0.04, than we 
calculate for the outer region.  The source of this discrepancy is 
unclear, but is possibly due to a difference in photometric depth. As we 
discussed in \S 4.3, reaching fainter features on the RGB LF helps to 
ameliorate the degeneracy between age, metallicity, and reddening.  Our 
outer region LF (Fig.~\ref{lf_err}) shows that we barely reach the RC/HB 
in NGC 1569.  \citet{mcquinnetal2010}, on the other hand, use only a 
subset of our data.  A comparison of their published CMD of the entire 
NGC 1569 field (their Fig.~3) with our CMD of the core (see 
Fig.~\ref{core_cmd}) shows that our photometry reaches 0.5 - 0.75 mag 
deeper.  Thus, their calculation of the reddening and distance does not 
benefit from the added constraints provided by the RC/HB. It is 
interesting to note, however, that even though \citet{mcquinnetal2010} 
do not reach the same photometric depth as we do, their SFH is very 
similar to ours.  This suggests that even without the RC/HB feature a 
significant amount of information can still be gleaned from the CMD.

\section{Possible Star Formation Triggers}
\label{triggers}

The central region of NGC 1569 is currently undergoing a strong 
starburst. Our CMD analysis of its outer region indicates that a 
constant level 
of star formation has been maintained there for much of the Hubble time. 
This prompts the question of what may have triggered this star formation 
activity.
 
Its projection on the sky places NGC 1569 near the IC 342 group.  The IC 
342 group is comprised of nine known galaxies and has a mean 
distance of $3.35\pm0.09$ Mpc, with a line-of-sight depth of 0.25 Mpc 
($1\sigma$; \citealt{karachentsev2005}). NGC 1569 was long assumed to 
lie only 2.2 Mpc away (e.g., \citealt{israel1988}), placing it in a 
relatively isolated position between the Local Group and the IC 342 
group. However, the TRGB distance inferred from our HST data, $D = 3.06 
\pm 0.18$ Mpc, places NGC 1569 inside the IC 342 group, raising the 
possibility that interactions with the group or its galaxies may have 
triggered the star formation in NGC 1569 \citep{grocholskietal2008}.

In the rest frame of the Local Group, the IC 342 group (excluding NGC
1569 and UGCA 92, see below) has a line-of-sight velocity $\langle
V_{LG} \rangle$ = 226$\pm$18 km s$^{-1}$, with a dispersion of 54 km
s$^{-1}$ (\citealt{karachentsev2005}). NGC 1569 has a lower velocity,
$V_{LG}$ = 88 km s$^{-1}$, that is 2.5$\sigma$ from the group
average. So NGC 1569 may reside in the tail of the velocity
distribution of the IC 342 group, or it may not be bound to it at all
(depending on its unknown transverse velocity, $v_{\rm trans}$).

Since NGC 1569 lies at the front side of the IC 342 group, NGC 1569 is 
now moving away from it. Given the known line-of-sight velocities and 
distances, we can calculate that it was at the same distance as the IC 
342 group $\sim 2.1$ Gyr ago, and entered the far side of the IC 342 
group $\sim 3.8$ Gyr ago. Fig.~\ref{sfrvage} shows that the NGC 1569 
outer region had a relatively constant non-zero star formation rate in 
this 
period. Thus, interactions with the IC 342 group, whether from tidal 
forces between galaxies or through ram pressure compression by IGM gas 
may have played a role in its SFH.

It is possible that NGC 1569 may have interacted strongly with a 
particular galaxy in the IC 342 group. One candidate is the dwarf galaxy 
UGCA 92. Like NGC 1569, UGCA 92 has a distance of $\sim$ 3 Mpc and 
$V_{LG} = 89$ km s$^{-1}$ \citep{karachentsevetal2006}.  They are 
separated on the sky by only $1\fdg23$, which gives a physical distance 
of $\sim$ 65 kpc.  A recent study by \citet{jacksonetal2011} shows that 
NGC 1569 is sitting in a large, cold H{\small I} cloud, with tidal tails 
stretching out toward UGCA 92. The combination of small spatial 
separation, similar line-of-sight velocity, and the H{\small I} gas 
possibly connecting the two galaxies suggests that NGC 1569 and UGCA 92 
may be interacting with each other. It is therefore possible that tidal 
forces between these galaxies have driven the long term SFH of both 
galaxies.  Dynamical models have shown that close encounters between two 
dwarf galaxies, such as the Magellanic Clouds, can lead to widespread 
star formation (e.g., \citealt{bekkietal2004}; Besla \etal in prep.). 
However, the distance between NGC 1569 and UGCA 92 is about three times 
the current separation between the Magellanic Clouds.  In absence of 
knowledge about $v_{\rm trans}$, it is not possible to know whether NGC 
1569 may have interacted directly with IC~342 itself. Also, the distance 
to IC~342 is not particularly well known, with many widely separated 
values quoted in the literature (e.g., 
\citealt{tikhonovandgalazutdinova2010, sahaetal2002} and references 
therein).

\section{Summary and Conclusions}
\label{summary}

NGC 1569 is one of the closest and strongest starburst galaxies in the
Universe. While its ISM and recent star formation have been thoroughly
studied, until recently, little was know about NGC 1569's old stellar
populations. This has made it difficult to determine the exact
duration of the starburst, to unravel its triggers, and to understand
NGC 1569's evolution over a full Hubble time. For this reason, we used
HST/ACS to obtain deep $V$- and $I$-band images that reach down to
$M_V \sim$ -0.5. These data allowed us to show definitively for the
first time that NGC 1569 has a significant population of stars older
than 1 Gyr \citep{grocholskietal2008}. Here we have used the same data
to derive the properties and full SFH of this old population.

We focused our analysis on the outer region of the galaxy, 
which is largely devoid of young stars and is much less crowded than the 
core. The improved photometric depth provides access to the RC/HB, which 
helps to ameliorate the age-metallicity degeneracy that plagues SFH 
studies that focus only on the RGB. We have used first an approach based 
on simple stellar populations. We then presented a newly developed 
synthetic CMD fitting code, SFHMATRIX, and used it to determine the full 
SFH. By combining the results from the different approaches, we were 
able to derive the following coherent picture for NGC 1569's old stellar 
populations.

{\bf (1)} By treating NGC 1569's outer region as an SSP, we find that 
the 
relative brightnesses of observed LF features (TRGB, RC/HB, and AGB 
bump) are best matched by a SSP with age = 10 Gyr and [Fe/H] = -1.4. 
However, the model SSP LF slope is flatter and the features more 
pronounced that what we observe in NGC 1569. This discrepancy suggests 
that NGC 1569 cannot be treated as an SSP as it must have had a more 
complex SFH.

{\bf (2)} The distance/reddening combination that best fits the data 
with a full SFH analysis is \ebv = 0.58 $\pm$ 0.03 and $D = 3.06 \pm 
0.18$ Mpc.  The reddening is in agreement with values published using 
other techniques, but the distance is $\sim$50\% farther than what has 
been typically assumed.

{\bf (3)} Star formation in the outer region of NGC 1569 began $\sim$ 13 
Gyr
ago and lasted until $\sim$ 0.5 Gyr ago.  The majority of star
formation in our observed field occurred early on, with 93\% of the
stars, by mass, having formed more than 5 Gyr ago.  This initial burst
was followed by a relatively low, but constant, rate of star formation
until $\sim$ 0.5-0.7 Gyr ago when there may have been a short, low
intensity burst of star formation.

{\bf (4)} The SFH for the old population in the NGC 1569 outer region 
follows
a trend of increasing metallicity with time. The youngest significant
population of $\sim$ 0.5-0.7 Gyr age has a metallicity similar to that
of the ionized gas in NGC 1569. These results are consistent with the
basic expectations of chemical evolution scenarios.

{\bf (5)} NGC 1569's dominant old population (age $\ga$ 10 Gyr) shows a 
considerable spread in metallicity, ranging from [Fe/H] = -1 to -2, with 
a peak around [Fe/H] = -1.25.  The mean of this distribution is very 
similar to what we derived in our SSP analysis.  The metallicity-spread 
of the dominant old population is similar to that for the Milky Way 
halo. However, the star formation in NGC 1569's outer region extended 
for much 
longer than in the Milky Way, indicating that NGC 1569 likely had 
atypical environmental influences.

{\bf (6)} The distance and line-of-sight velocity of NGC 1569 indicate
that it moved through the IC 342 group of galaxies, at least in the
past few Gyr. This may be the reason for the extended low-level star
formation seen in its outer region.  By contrast, its recent starburst 
may be
related to interactions with the companion UGCA 92.

{\bf (7)} Comparison with recent work from \citet{mcquinnetal2010} and 
\citet{rysetal2011}, which was more heavily weighted towards smaller and 
larger radii in the galaxy, respectively, provides no evidence for 
radial population gradients in the old population of NGC 1569. This 
suggests that our results for the outer region are representative for 
the old 
stellar population throughout the galaxy.
 
\acknowledgments

We would like to thank Jay Anderson for helpful discussions regarding 
early versions of the photometry, Leo Girardi for answering many 
questions about the Padova models, and Kristen McQuinn for kindly 
providing an electronic version of her SFH for NGC 1569.  The authors 
would also like to thank the anonymous referee for comments that helped 
to improve the clarity of the paper.  We are grateful to Jennifer Mack, 
Luca Angeretti, Enrico Held, Donatella Romano and Marco Sirianni for 
collaboration on earlier stages of this project. Support for proposal 
GO-10885 was provided by NASA through a grant from STScI, which is 
operated by AURA, Inc., under NASA contract NAS 5-26555.  F.A. and M.T. 
acknowledge partial financial support from ASI through contracts 
ASI-INAF I/016/07/0 and I/009/10/0.\\

{\it Facility:} \facility{HST (ACS)}.

\appendix

\section{Star Formation History Analysis with SFHMATRIX}

The Star Formation History (SFH) of a galaxy can be described
mathematically as a sum of delta functions. Each delta function has a
fixed combination of age and metallicity, and a weight corresponding
to the mass of stars that formed with that combination. This implies
that the CMD of a galaxy can be described as the weighted linear sum
of a set of ``basis functions'', where each basis function corresponds
to the synthetic CMD for an isochrone with fixed age and metallicity
\citep{tosietal1991}. The basis functions must
reflect the characteristics of the problem at hand, namely: (1)
observational details, such as the completeness and photometric errors
inferred from artificial star tests; (2) galaxy properties, such as
the distance and total (foreground+internal) extinction; and (3)
stellar population properties, such as the Initial Mass Function (IMF)
with which an isochrone is populated, and the binary fraction.

Framed in this manner, the problem of inferring the SFH of a galaxy
from its CMD reduces to the problem of finding the weighted
combination of synthetic basis function CMDs that best reproduces the
observed CMD. Since the data consist of a discrete set of points, this
can be expressed mathematically as a maximum likelihood problem (e.g., 
\citealt{dolphin2002}). However, when the number of basis
functions is large, such problems can be complicated to solve
numerically. It is therefore advantageous to consider instead the
density of points in the CMD, i.e., the Hess diagram, rather than the
CMD itself (both for the observations and the basis functions). By
pixelating the CMD and assigning an appropriate error bar to the
observed density in each pixel, the problem reduces to a linear
$\chi^2$ minimization problem (e.g., \citealt{harrisandzaritsky2001}). 
For large numbers of stars, these approaches
become mathematically equivalent due to the central limit theorem.

Many software implementations have been developed to infer the SFH
from an observed CMD, mostly using approaches similar to those
outlined above (see, e.g., the compilation and comparison of various
methods described in \citealt{skillmanandgallart2002}).
Here we use a new code that we developed, called SFHMATRIX. We
describe here the salient features, with a more detail description
planned for a separate paper (van der Marel \& Grocholski in
prep.). Our code resembles the STARFISH code of 
\citet{harrisandzaritsky2001}. 
Our method of constructing a synthetic CMD from an isochrone
is largely similar to theirs, and like STARFISH, we describe the
problem mathematically as a $\chi^2$ minimization.

Our code differs primarily from the STARFISH implementation in how it
finds the $\chi^2$ minimum, i.e., the best-fitting SFH. STARFISH
finds the minimum using a brute-force minimization in $N$-dimensional
space, where $N$ is the number of basis functions (i.e., the number of
different isochrone weights to be optimized). By contrast, we phrase
the problem as the solution of a matrix equation (hence the name
``SFHMATRIX'')
\begin{equation}
  \sum_{j=1}^N A_{ij} m_j = \rho_i \pm \Delta \rho_i , \qquad 
  \forall i=1,\ldots,M \qquad .
\end{equation} 
Here $\rho_i$ is the density of the stars in CMD space, $\Delta
\rho_i$ is the associated Poisson uncertainty, and the index $i$
counts the $M$ pixels of the Hess diagram. 
The density $\rho_i$ equals the integer number of stars $L_i \geq 0$ 
that is detected in Hess diagram pixel $i$, divided by the area of that 
pixel. We adopt $max(1,\sqrt{L_i})$ as the Poisson error on the detected 
number of stars. Hence, $\Delta \rho_i / \rho_i = max(1,\sqrt{L_i}) / 
L_i$.
The vector of weights $m_j \geq 0$ gives the
mass (in $M_{\odot}$) of stars associated with the $N$ basis functions
(i.e., isochrones) $j$. The matrix $A_{ij}$ gives the density of stars
in Hess diagram pixel $i$ for a $1 M_{\odot}$ population of stars on
synthetic isochrone $j$. Finding the solution of this matrix equation
is a non-negative least squares (NNLS) problem in linear algebra. This
can be solved with efficient general purpose subroutines (e.g., 
\citealt{lawsonandhanson1974}) that have been well tested in other areas 
of astronomy (e.g., the dynamical modeling of galaxies using
Schwarzschild's orbit superposition technique; e.g., 
\citealt{vandermareletal1998}).

Our approach has several advantages over the STARFISH approach. First,
the NNLS matrix routines are guaranteed to converge on a global
minimum \citep{lawsonandhanson1974}, and cannot inadvertently find local
minima as do brute-force searching routines. Second, the speed of NNLS
routines makes it easy to use many basis function (large $N$), so that
it is possible to effectively search and characterize the space of
relevant models on a fine grid (both in age and metallicity, as well
as in other parameters such as extinction and distance). Third, the
speed of NNLS routines makes it easy to run the code in Monte-Carlo
sense on many simulated datasets, so that the uncertainty on the final
SFH can be robustly characterized.

Our code can create basis functions from any set of isochrones. It is
important that the basis functions fully span the range of ages and
metallicities that are relevant for the galaxy under study. It is also
important that the spacing between adjacent isochrones is relatively
fine. If it is too coarse, then a fitted CMD superposition looks
choppy and discontinuous in those regions of CMD space where the
observational errors are small. On the other hand, the computational
effort of the NNLS solution scales with the number of basis functions
as $N^3$. So it is best not to choose more basis functions than can
realistically be resolved by the data given the observational errors.

Here we use the isochrones from the same Padova models as discussed in 
\S\ref{LFs} and we note that these are the same isochrones used in 
the paper by \citet{mcquinnetal2010}.  We use metallicities ranging 
from -2.28 dex to +0.20 dex and log(age) values from 6.60 to 10.12. We 
obtained isochrones spaced by 0.05 dex in metallicity and 0.01 dex in 
log(age). Most of these isochrones are obtained from interpolation 
between a much coarser set of isochrones (in particular in metallicity) 
for which actual stellar evolutionary calculations were performed (e.g., 
\citealt{fagottoetal1994a,fagottoetal1994b}). We created synthetic CMDs 
from each basis function by randomly drawing many stars from the 
isochrone, given an assumed IMF, and then rescaling the corresponding 
Hess diagram density to correspond to $1 M_{\odot}$. Stars are assigned 
a simulated photometric error and are marked as either detected or not 
(thus accounting for incompleteness) based on a randomly drawn 
artificial star from the artificial star test results. To construct 
basis functions that are not sampled quite as finely, we co-add the 
synthetic CMD results within bins of size 0.1 dex in metallicity and 0.1 
dex in log(age).\footnote{This results in each basis function 
corresponding to a two-dimensional comb function that approximates a 
two-dimensional boxcar. The resulting SFH is therefore more akin to a 
two-dimensional histogram, than a sum of delta-functions.}

Finding the best-fitting SFH from an observed CMD is mathematically an
inverse problem. Methods for solving such problems have the well-known
tendency for amplifying noise, leading to solutions that appear
unphysically spiky. This can be counteracted by enforcing smoothness
on the solution. Regularization is one popular technique for doing
this (e.g., \citealt{pressetal1992}).  In the context of NNLS solutions,
linear regularization constraints can be enforced by adding $K$
additional rows of the following form to the matrix equation
\begin{equation}
  \sum_{j=1}^N B_{kj} m_j = 0 \pm \Delta , \qquad 
  \forall k=1,\ldots,K \qquad .
\end{equation}
(see \citealt{crettonetal1999}). The $B_{kj}$ are chosen
such that for given $k$, the expression $\sum B_{kj} m_j$ is a second
order divided difference of basis functions that are adjacent in age
at fixed metallicity, or vice versa. The divided differences are zero
if the SFH is locally well approximated by a linear function. In this
manner, smoothness is enforced in (age,metallicity) space, with the
exact amount depending on the parameter $\Delta$. If $\Delta =
\infty$, then the regularization constraints are ignored. If $\Delta =
0$, then the data are ignored, and the code cares only about returning
a smooth solution. For intermediate values, the NNLS solutions tries
to fit the data as best as possible, while keeping the noise in the
SFH in check. In practice we applied a small amount of regularization
to obtain the results in this paper, but none of our main results
depend sensitively on this.

Calculating error bars on the best-fitting SFH returned by the code is
straightforward. For this we create many realizations of pseudo
datasets with properties similar to the real data. We analyze each of
these pseudo datasets in Monte-Carlo fashion as we do the real
data. The RMS scatter in the SFH results at a given (age,metallicity)
combination is the error bar. The pseudo data can be created with
either of two possible approaches. The first approach is to use
bootstrapping \citep{pressetal1992}. In this approach one obtains a new
dataset of $S$ stars from the existing dataset of $S$ stars, by random
drawing with replacement. The second approach is to use the
best-fitting SFH already inferred from the real data, and to draw many
Monte-Carlo pseudo dataset realizations from this SFH. In the present
paper we use the bootstrapping approach to calculate SFH error bars.
 
To test the accuracy of the code we explored two approaches. In the
first approach we used pseudo dataset realizations drawn from a known
SFH, and then used the code to verify that the inferred SFH agrees
with the input SFH to within the errors. In the second approach we
used our code and the publicly available STARFISH code on the same
input data, and verified that the inferred SFHs agreed to within the
errors. Both tests were passed successfully. 

We also explored the sensitivity of the results to the small changes in 
the stellar evolutionary models using two approaches. In the first 
approach we used our code with either the isochrones from 
\citet{bertellietal1994} or \citet{girardietal2002} to infer the SFH 
from the same data. In the second approach we used our code with basis 
functions that were created either from the \citet{bertellietal1994} 
isochrones, or directly from the 
\citet{fagottoetal1994a,fagottoetal1994b} evolutionary tracks from which 
these isochrones were derived, and then derived the SFH from the same 
data. The approaches all yielded results in satisfactory agreement. 
However, more generally this does depend on both the actual differences 
in the underlying evolutionary calculations, and the characteristics and 
quality of the available data.  Different stellar evolutionary 
tracks can yield different implied SFHs when used to analyze the same 
data (e.g., Skillman \& Gallart 2002).

We have assumed a Salpeter IMF \citep{salpeter1955} from $0.1-100$ 
M$_{\odot}$ in the calculations presented in this paper, similar to 
prior studies of the SFH in NGC 1569 
(\citealt{mcquinnetal2010,angerettietal2005}). The exact IMF choice does 
not affect the shape of the inferred SFH for the outer region of NGC 
1569. This 
is because all observed stars have evolved off the main sequence, and 
have approximately the same mass. However, the IMF choice does affect 
the normalization of the SFH, since different IMF slopes predict 
different amounts of (unobserved) lower-mass stars for a given 
(observed) population of giants. The assumed binary fraction also 
affects the SFH normalization, because some single observed sources may 
have the mass of two stars. For the calculations in the paper the SFH 
shape was not found to depend sensitively on the assumed binary 
fraction, and all results reported herein assume a zero binary fraction. 
Our code has the option to assign different reddening to different stars 
in the basis functions. We did not use this feature for the outer region 
of NGC 
1569. Our H$\alpha$ image showed very little emission there, so that 
differential reddening is not expected to be an issue.

\bibliography{biblio}

\end{document}